\documentclass[final,10pt]{elsarticle}

\usepackage[utf8]{inputenc}
\usepackage[T1]{fontenc}
\usepackage{mathptmx}
\usepackage[margin=1in]{geometry}  % Reduce margins
\usepackage{hyperref}
\usepackage{url}
\usepackage{booktabs}
\usepackage{amsfonts}
\usepackage{nicefrac}
\usepackage{microtype}
\usepackage{xcolor}
\usepackage{graphicx}
\usepackage{subcaption}
\usepackage{multirow}
\usepackage{array}
\usepackage{tabularx}
\newcolumntype{L}{>{\raggedright\arraybackslash}X}
\usepackage{makecell}
\usepackage{amsmath}
\usepackage{amssymb}
\usepackage{algorithm}
\usepackage{algorithmic}

\begin{document}

\begin{frontmatter}

\title{CrisiSense-RAG: Crisis Sensing Multimodal Retrieval-Augmented Generation for Rapid Disaster Impact Assessment}

\author[1]{Yiming Xiao\corref{cor1}}
\ead{yxiao@tamu.edu}

\author[2]{Kai Yin}
\ead{kai_yin@tamu.edu}

\author[1,2]{Ali Mostafavi}
\ead{amostafavi@civil.tamu.edu}

\cortext[cor1]{Corresponding author}

\address[1]{Zachry Department of Civil and Environmental Engineering, Texas A\&M University}
\address[2]{Department of Computer Science and Engineering, Texas A\&M University}

\begin{abstract}
    Timely and spatially resolved disaster impact assessment is essential for effective emergency response. However, automated methods typically struggle with temporal asynchrony. Real-time human reports capture peak hazard conditions, while high-resolution satellite imagery is frequently acquired after peak conditions. This often reflects flood recession rather than the maximum extent. Naive fusion of these misaligned streams can yield dangerous underestimates when post-event imagery overrides documented peak flooding. We present CrisiSense-RAG, a multimodal retrieval-augmented generation framework that reframes impact assessment as evidence synthesis over heterogeneous data sources without disaster-specific fine-tuning. The system employs hybrid dense-sparse retrieval for text sources and CLIP-based retrieval for aerial imagery. A split-pipeline architecture feeds into asynchronous fusion logic that prioritizes real-time social evidence for peak flood extent while treating imagery as persistent evidence of structural damage. Evaluated on Hurricane Harvey across 110 ZIP-code queries with multimodal coverage, the full multimodal configuration achieves a flood extent MAE of 8.86\% to 19.64\% and damage severity MAE of 10.14\% to 14.47\% across four model backends. Full multimodal RAG reduces flood extent MAE by up to 8.62 percentage points relative to text-only inference. Prompt-level alignment proves critical for quantitative validity: without explicit metric grounding, models interpret damage severity as qualitative intensity rather than the statistical average fraction of building value destroyed. These results demonstrate a practical and deployable approach to rapid resilience intelligence under real-world data constraints.
\end{abstract}


\end{frontmatter}

%==============================================================================
% INTRODUCTION
%==============================================================================
\section{Introduction}\label{sec:introduction}

Rapid disaster impact assessment is a time-critical prerequisite for effective emergency management. Responders must quickly determine both where a hazard is occurring (e.g., flood extent) and how severe its consequences are (e.g., structural damage) to prioritize rescues and allocate scarce resources. Yet, dominant approaches remain fundamentally constrained: manual ground surveys are slow and dangerous, while physics-based hydrodynamic models are computationally expensive and struggle to reflect fast-evolving conditions at operational tempo~\citep{bryan-smith_real-time_2023}. As climate-driven extremes increase, there is an urgent need for scalable methods capable of deriving timely impact estimates from the data streams already generated during disasters~\citep{liu_artificial_2025, krichen_managing_2024}. Among available spatial units, the ZIP code strikes a practical balance for decision support: it is fine-grained enough to differentiate neighborhood-level impacts yet coarse enough to aggregate noisy, heterogeneous reports into statistically meaningful estimates, and it aligns directly with the administrative geographies through which Federal Emergency Management Agency (FEMA) aid is disbursed and local resources are allocated~\citep{fema_ia_program_2019, cutter_social_2003}. The past decade has brought a proliferation of potentially informative data, from satellite imagery to social media streams. However, these sources are not merely heterogeneous in format; they are asynchronous in timing. Human reports and emergency calls typically peak during the hazard (capturing rising waters), while aerial and satellite imagery is frequently collected hours or days later (capturing recession).

This temporal misalignment presents a critical challenge for existing automated assessment methods. Much of the literature relies on modality-specific modeling or ``naive fusion'' approaches that aggregate features into a single vector, such as computer vision for flood mapping or natural language processing (NLP) for situational awareness~\citep{yuan_smart_2022, dong_hybrid_2021}. These strategies implicitly treat each data stream as a static, contemporaneous signal. When applied to asynchronous disaster data, this creates a dangerous mismatch: post-event clear-sky imagery may statistically negate earlier high-confidence flooding reports, producing underestimates precisely when accurate situational awareness is most critical. Furthermore, conventional supervised multimodal learning remains difficult to operationalize because it depends on large, labeled, multi-event datasets that are rarely available for emerging platforms or novel disaster scenarios.

To address these limitations, we reframe impact assessment as an evidence synthesis problem rather than a single-modality prediction task. We propose a Multimodal Retrieval-Augmented Generation (RAG) framework that leverages pre-trained foundation models to synthesize diverse data streams without event-specific fine-tuning. Unlike traditional fusion, our architecture explicitly reconciles temporal misalignment through a split-pipeline design. We employ separate Text and Visual Analyst modules to reason about modality-specific evidence, joined by an asynchronous fusion logic that treats real-time social data as evidence for peak hazard extent, while treating post-event imagery as persistent evidence of structural damage. Crucially, we enforce metric-aligned generation, ensuring outputs conform to standardized operational definitions (e.g., average damage per building, peak inundation depth) rather than qualitative severity language.

The central objective of this study is to determine whether general-purpose foundation models can deliver spatially resolved, quantitative impact estimates in a zero-shot setting.\footnote{By ``zero-shot'' we mean that no event-specific fine-tuning was performed on any large language model (LLM) weights. The system does use domain-general disaster-response prompts and FEMA historical priors, but the model parameters themselves are used as released.} We validate this approach using Hurricane Harvey, a well-documented event with diverse data availability and measurable ground truth. Our contributions are summarized as follows:

\medskip
\textbf{(1) Temporally-asynchronous fusion}
\medskip

We propose a split-pipeline Multimodal RAG architecture with temporal-aware fusion logic that explicitly reconciles the gap between real-time text signals (capturing peak hazard) and post-event imagery (capturing recession). Unlike existing multimodal RAG systems that assume contemporaneous inputs, our fusion rules prevent post-event imagery from negating documented peak flooding, addressing a failure mode absent from prior work.

\medskip
\textbf{(2) Heterogeneous five-source integration}
\medskip

 We demonstrate a single retrieval-to-generation pipeline that unifies five structurally different modalities (social media, institutional emergency calls, precipitation data, aerial imagery, and historical loss priors) under hybrid dense-sparse retrieval with cross-encoder reranking and CLIP-based visual search, without requiring modality-specific supervised training.

\medskip
\textbf{(3) Evidence-grounded output}
\medskip

 Every quantitative estimate (flood extent, damage severity) is traceable to specific document IDs (\texttt{evidence\_refs}: tweet IDs, 311 call IDs, imagery tile IDs, sensor IDs), enabling emergency responders to verify and challenge predictions. This level of traceability is absent from prior multimodal fusion systems.

\medskip
\textbf{(4) Rigorous multi-model evaluation}
\medskip

 We evaluate on 110 ZIP-code queries with imagery coverage across four diverse model backends (Gemini 2.5 Flash~\citep{gemini2025}, Gemini 3 Flash~\citep{gemini2025}, Qwen 3.5-397B-A17B~\citep{qwen2025} (referred to as Qwen 3.5 in the following text), and GPT-5-mini~\citep{openai2025gpt5}), showing that the full multimodal configuration attains 8.86--19.64\% Extent MAE and 10.14--14.47\% Damage MAE across backends. Full multimodal RAG reduces flood extent MAE by up to 8.62 percentage points relative to text-only inference, demonstrating that aerial imagery provides complementary evidence for hazard estimation.

%==============================================================================
% RELATED WORK
%==============================================================================
\section{Related Work}
\label{sec:related_work}

\subsection{Remote Sensing and Automated Damage Assessment}
Satellite imagery is the standard for large-scale disaster monitoring. Recent research has focused on Deep Learning (DL) models to automate the extraction of damage features and post-event change signals at scale. \citet{kaur_largescale_2023} introduced hierarchical transformers (DAHiTrA) for building localization and graded damage classification in large benchmark settings, highlighting the effectiveness of multi-resolution representations for post-disaster mapping. Similarly, \citet{zarski_multi-step_2024} proposed multi-step feature fusion networks to improve change detection between pre- and post-event imagery, reducing missed detections when changes are subtle. To address the limitations of binary or ordinal severity scales, \citet{xiao_damagecat_2025} introduced DamageCAT, a typology-based framework that classifies damage into actionable categories (partial roof damage, total collapse, etc.) using hierarchical transformers on satellite image triplets, demonstrating improved transferability across multiple hurricane events. In a different direction, \citet{pandey_google_2022} demonstrated the utility of Synthetic Aperture Radar (SAR) for flood mapping in cloud-obscured regions, emphasizing the operational value of all-weather sensing during active storm systems. For building-level flood damage, \citet{ho_multimodal_2025} proposed a risk-aware multimodal architecture that fuses SAR/Interferometric SAR (InSAR) with optical basemaps to jointly estimate floodwater extent and graded damage states, explicitly targeting cases where flood damage is nonstructural and difficult to observe in imagery alone. More recently, \citet{ho_integrated_2025} integrated street-view imagery with vision-language models to estimate building elevations, illustrating how street-level visual context can complement overhead sensing for exposure-related attributes. Extending this street-level paradigm to post-disaster recovery, \citet{xiao2025recovvisionlinkingstreetview} proposed Recov-Vision, a framework that links panoramic street-view video to building parcels using vision-language models to assess occupancy and habitability, capturing facade-level cues (entry blockage, temporary coverings) that overhead imagery cannot observe.

A parallel line of work has focused on large-scale labeled benchmarks to enable supervised building-level assessment. \citet{gupta_xbd_2019} introduced xBD, the largest building damage dataset to date, with 850,000 annotations across 45,000 km$^2$ of pre/post satellite imagery spanning multiple disaster types, enabling ordinal damage classification at the individual parcel level. \citet{rahnemoonfar_floodnet_2021} released FloodNet, a high-resolution unmanned aerial vehicle (UAV) imagery dataset captured after Hurricane Harvey specifically, supporting semantic segmentation and visual question answering for post-flood scene understanding. These supervised approaches achieve fine-grained, building-level outputs but require pre/post image pairs, labeled training data, and cloud-free satellite revisits. None of them is reliably available in the immediate hours after landfall. Furthermore, these remote sensing methods remain limited by satellite revisit times (latency) and viewing angles, often struggling to capture lateral damage or interior flooding without corroborating ground-level evidence. CrisiSense-RAG targets a complementary granularity: ZIP-code-level situational awareness from heterogeneous real-time signals, aligned with how emergency managers actually allocate staging resources, trigger evacuation zones, and activate FEMA NFIP policies.

\subsection{Social Media for Disaster Situational Awareness}
Social media has emerged as a critical source of real-time disaster intelligence, providing a ``human sensor'' network that complements remote sensing through rapid, ground-level observations. \citet{acikara_contribution_2023} reviewed the field, organizing common social analytics tasks around location awareness and situational awareness, and noting persistent challenges in noise, bias, and verification; more broadly, \citet{zhang_social_2019} surveyed how social media supports public information and warning, peer-to-peer help, and agency listening during disasters. Specific applications include real-time sentiment analysis for flood impact~\citep{bryan-smith_real-time_2023}, where online discourse is used as a proxy signal for evolving impacts, and multimodal classification of disaster losses~\citep{zhou_classification_2025}, which targets fine-grained loss categories and severity cues from mixed text-image posts. Departing from conventional classifiers, instruction fine-tuning has been explored to adapt general-purpose LLMs for disaster-specific, multi-label tweet classification~\citep{yin2025crisissensellminstructionfinetunedlarge}, enabling one model to extract multiple actionable labels from a single post. \citet{zhang_semiautomated_2020} developed the SocialDISC framework to sense downstream societal impacts (e.g., power outages, road closures) from Twitter, illustrating how social streams can be transformed into structured operational indicators.

However, reliability remains a major bottleneck. \citet{mandal_algorithmic_2024} conducted a critical study on geolocation uncertainty, finding that less than 1\% of tweets are geotagged and that user profile locations have significant error margins, which can confound place-based analysis. In addition to location, the credibility of user-generated content is a persistent challenge, with noise, rumors, and hyperbole necessitating robust verification strategies. \citet{sathianarayanan_extracting_2024} attempted to mitigate location uncertainty by extracting visual cues (like phone numbers) from images, offering an alternative pathway to recover location when metadata is missing. Our work addresses these dual challenges of location and credibility by using RAG to retrieve semantically relevant reports and cross-referencing them with physical imagery.

\subsection{Multimodal Data Fusion and Resilience Frameworks}
The integration of heterogeneous data is essential for Smart Resilience, typically framed through the lens of hazard, exposure, and vulnerability, and operationalized via multi-source sensing and analytics. \citet{fan_disaster_2021} articulated a complementary vision of ``Disaster City Digital Twins'' that integrates artificial and human intelligence across heterogeneous data streams for situation assessment, decision support, and stakeholder coordination. Building on this vision, \citet{li2025disastermanagementeraagentic} proposed Disaster Copilot, a multi-agent AI architecture where specialized sub-agents coordinate through a central orchestrator to unify predictive analytics, situational awareness, and impact assessment into a real-time operational picture, addressing the fragmented data streams and siloed technologies that hinder current practices. \citet{yuan_smart_2022} proposed a framework fusing community-scale big data (traffic, social, credit card transactions) to monitor predictive flood risk, demonstrating how diverse proxy signals can track community-level disruption. \citet{dong_hybrid_2021} combined channel sensor data with deep learning for hybrid flood warning systems, illustrating how learned models can complement physical sensing for early warning. \citet{liu_flooddamagecast_2024} introduced FloodDamageCast, a machine learning framework for nowcasting property damage using built-environment features and data augmentation, underscoring progress toward timely consequence estimation beyond inundation mapping. While effective, these systems typically use early fusion (concatenating features) or separate model ensembles. They lack the semantic reasoning capabilities of Generative AI, which allows for the flexible interpretation of conflicting evidence (e.g., ``sensor says dry'' vs.\ ``human says help'') that is characteristic of complex, asynchronous disaster scenarios.

\subsection{Multimodal Retrieval-Augmented Generation (RAG)}
RAG has reshaped how LLMs incorporate external knowledge at inference time, reducing hallucinations and improving grounding when model weights alone are insufficient~\citep{oche_systematic_2025, wang_searching_2024}. The extension to multimodal RAG broadens this paradigm by enabling retrieval and reasoning over both text and images within a unified evidence pipeline. \citet{chen_murag_2022} pioneered this direction with MuRAG, which couples a transformer with non-parametric multimodal memory to retrieve and condition generation on relevant visual-text evidence. Recent surveys~\citep{abootorabi_ask_2025} document the rapid growth of this field, including design patterns for retrieval, indexing, and multimodal fusion in generation.
In domain-specific applications, RAG has shown promise in healthcare~\citep{zhu_realm_2024, xia_mmed-rag_2025}, where grounded generation is critical for safety and traceability, and in visual document understanding~\citep{luo_bi-vldoc_2025}, where retrieval helps connect layouts, text, and images to structured outputs. In the geospatial domain, \citet{chen_empowering_2025} demonstrated Geospatial Awareness Layers for LLM agents in wildfire response, highlighting how spatial context can improve tool-using reasoning, and \citet{yin_disastir_2025} introduced DisastIR, a benchmark that formalizes disaster information retrieval challenges for evaluation.

Table~\ref{tab:related_comparison} compares CrisiSense-RAG with representative prior systems. Existing multimodal RAG and vision-language model (VLM) pipelines either assume temporally aligned inputs, operate over fewer modality types, or produce outputs that lack per-prediction evidence traceability. Our work addresses all three gaps simultaneously: it explicitly handles temporally asynchronous evidence streams, integrates five structurally heterogeneous data sources under a single retrieval-to-generation pipeline, and provides document-level evidence references for every quantitative estimate.

% Comparison table
\begin{table}[!htbp]
    \centering
    \caption{Comparison of CrisiSense-RAG with representative multimodal RAG and disaster assessment systems across three key properties.}
    \label{tab:related_comparison}
    \begin{tabular}{lccc}
        \toprule
        \textbf{System} & \textbf{Temporal}$^{a}$ & \textbf{Modalities}& \textbf{Auditable}$^{b}$ \\
        \midrule
        MuRAG~\citep{chen_murag_2022}                      & --- & 2 & --- \\
        REALM~\citep{zhu_realm_2024}                       & --- & 2 & \checkmark \\
        MMed-RAG~\citep{xia_mmed-rag_2025}                 & --- & 2 & \checkmark \\
        GAL (Wildfire)~\citep{chen_empowering_2025}        & --- & 3 & --- \\
        Smart Resilience~\citep{yuan_smart_2022}           & --- & 4 & --- \\
        Disaster Copilot~\citep{li2025disastermanagementeraagentic} & --- & 3+ & --- \\
        \midrule
        \textbf{CrisiSense-RAG (Ours)}                    & \checkmark & 5 & \checkmark \\
        \bottomrule
        \multicolumn{4}{l}{\footnotesize $^{a}$Explicitly addresses asynchrony between data streams.} \\
        \multicolumn{4}{l}{\footnotesize $^{b}$Predictions traceable to specific source document IDs.} \\
    \end{tabular}
\end{table}

%==============================================================================
% METHODOLOGY
%==============================================================================
\section{Methodology}
\label{sec:methodology}

\begin{figure}[!htbp]
    \centering
    \includegraphics[width=\textwidth]{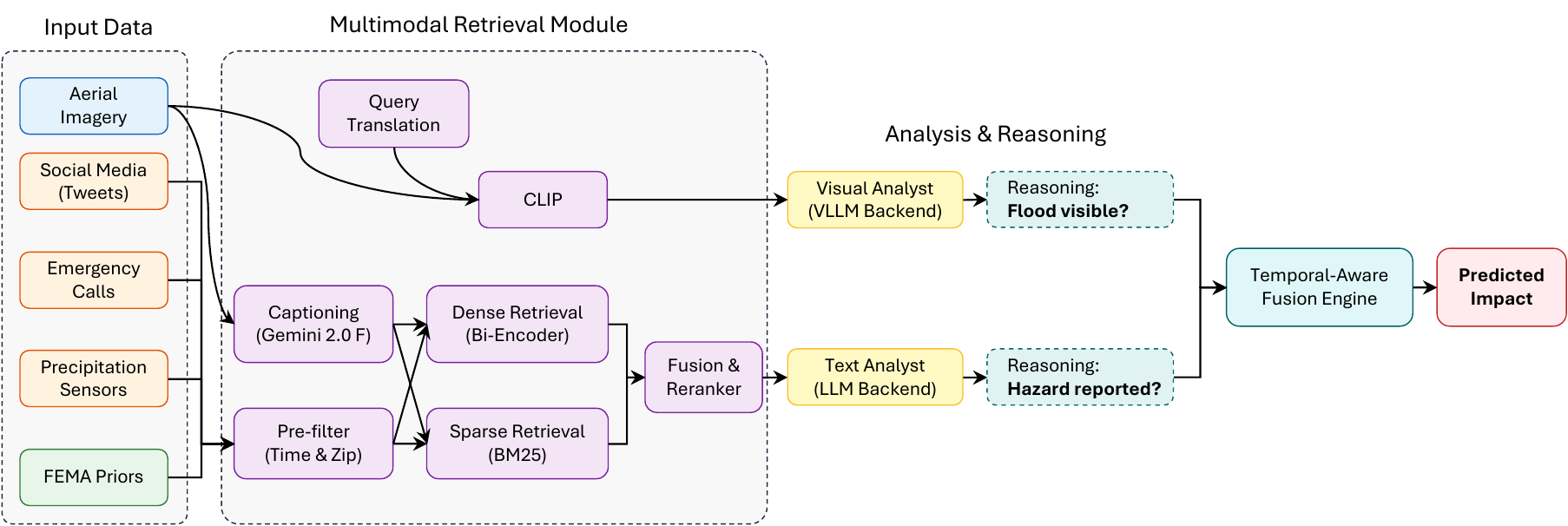}
    \caption{Overview of the proposed CrisiSense-RAG architecture.}
    \label{fig:system_architecture}
\end{figure}

We present a multimodal RAG system designed to estimate flood extent and structural damage from heterogeneous, asynchronous data streams. Figure~\ref{fig:system_architecture} illustrates our architecture, which integrates aerial imagery, social media posts, 311 emergency calls, precipitation data, and historical FEMA loss priors through a retrieval-augmented generation pipeline. The key design principle is a split-pipeline approach that processes text and visual modalities separately before fusion, respecting their distinct temporal characteristics.

\subsection{Task Definition}
\label{sec:task_definition}
We define the impact assessment task as estimating the peak flood extent and structural damage that occurred during the event (August 25--September 1, 2017), regardless of whether the water has receded at the time of retrieval. This formulation aligns with the operational needs of recovery agencies, who require a record of maximum impact rather than a transient snapshot of current conditions. Consequently, our system is designed to prioritize evidence of peak inundation (e.g., real-time social reports) over post-event evidence of recession (e.g., clear aerial imagery), unless permanent structural damage is visible.

\subsection{Data Sources and Preprocessing}

\label{sec:data_sources}
\begin{figure}[!htbp]
    \centering
    \includegraphics[width=0.6\textwidth]{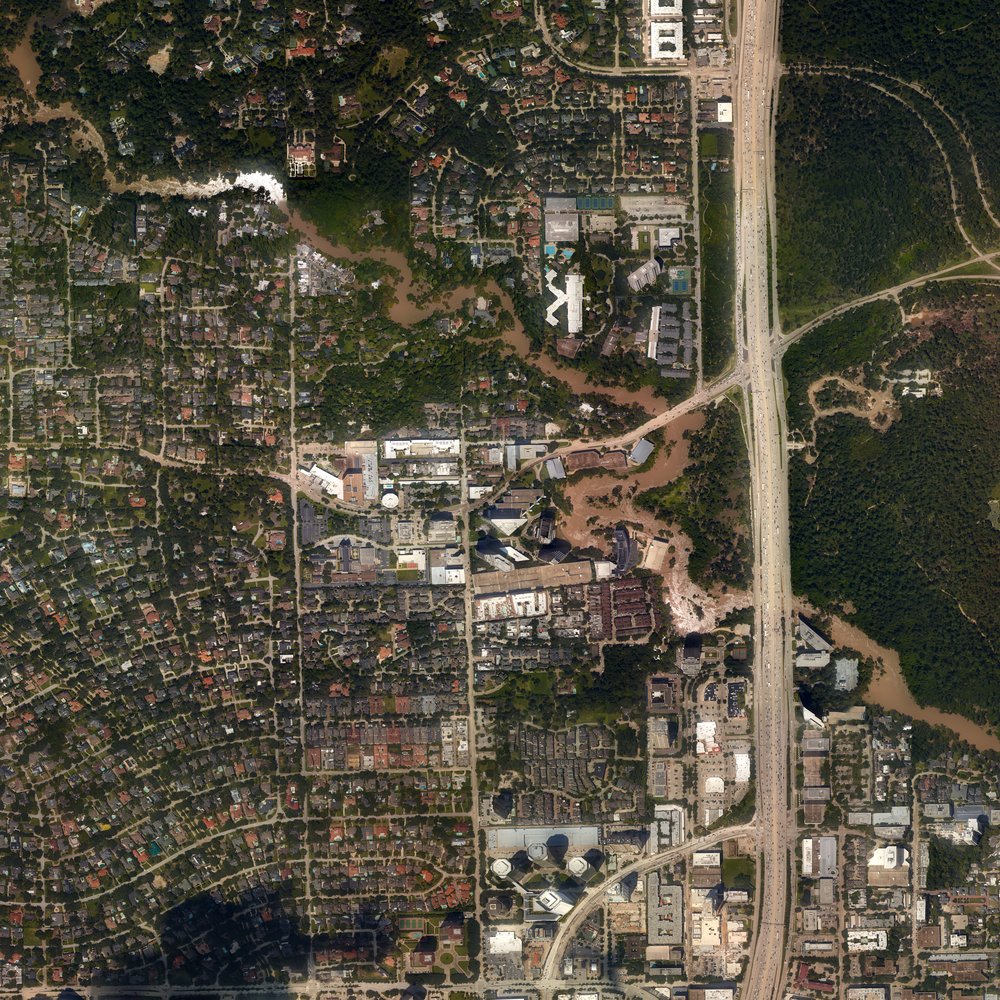}
    \caption{Example of a retrieved aerial imagery tile (Aug 31, 2017). The LLM-generated caption successfully identifies key features: \textit{``Significant flooding is visible, with brown water inundating areas along a winding waterway\ldots and encroaching on the highway in several places.''}}
    \label{fig:example_tile}
\end{figure}

Our system integrates five primary data sources collected during the event, each providing complementary perspectives on disaster impact.

\medskip
\textbf{(1) Aerial Imagery}
\medskip

We utilize high-resolution aerial imagery obtained from NOAA's Emergency Response Imagery program~\citep{noaa_ngs_imagery}. The dataset comprises 3,507 georeferenced tiles covering the Greater Houston metropolitan area at 0.5~m spatial resolution. Each tile is indexed by geographic coordinates and acquisition timestamp to enable spatiotemporal queries.

\medskip
\textbf{(2) FEMA Prior Knowledge}
\medskip

To provide historical context, we retrieve aggregated historical financial loss records from FEMA's Individual Assistance database (2010--2016). This includes the average verified loss amount for the ZIP code from prior declared disasters. Crucially, this data represents historical vulnerability profiles established years before Hurricane Harvey and is strictly disjoint from the 2017 event-specific ground truth (depth grids and damage estimates) used for evaluation, ensuring no data leakage.

\medskip
\textbf{(3) Social Media (Twitter/X)}
\medskip

We curated a corpus from approximately 24 million tweets collected during Hurricane Harvey, applying a multi-stage filtering pipeline to maximize relevance. First, we performed de-duplication by removing retweets, which constitute approximately 57\% of the raw data. Second, we applied content filtering using keyword-based allow and block lists: the allow list keeps tweets containing disaster-relevant terms (e.g., ``flood,'' ``rescue''), while the block list excludes off-topic content (e.g., ``spotify,'' ``music''). Third, we implemented spam removal to filter posts containing excessive hashtags or URLs. This pipeline yielded a final corpus of 458,453 unique tweets, representing an acceptance rate of approximately 1.9\%.

\medskip
\textbf{(4) 311 Emergency Calls}
\medskip

We obtain records from the City of Houston's 311 service system, including reports of flooding, debris, and infrastructure damage. The dataset contains 26,107 records from August 20--September 10, 2017, covering 125 unique ZIP codes in the urban Houston area.

\medskip
\textbf{(5) Precipitation Sensors}
\medskip

We incorporate data from Harris County Flood Control District rain gauges. We compute peak cumulative rainfall for each ZIP code by interpolating from the nearest gauge stations (centroid-based nearest neighbor lookup), providing a persistent, objective flood indicator as a spatial prior in the model prompt.

To enable semantic retrieval over aerial imagery, we additionally performed a preprocessing step in which Gemini 2.0 Flash~\citep{gemini2025} was used, for cost-efficiency, to generate natural language descriptions for all 3,507 aerial image tiles. These machine-generated captions are indexed as searchable text documents, enabling the retrieval system to surface relevant imagery based on textual queries describing observable damage patterns (e.g., submerged residential structures, debris accumulation). Thus, the captions function as retrieval metadata for the aerial imagery rather than as a separate data source.

\subsection{Multimodal Retrieval}
\label{sec:retrieval}

Given a query specifying a ZIP code and time window, our retrieval system identifies the most relevant context from each data source. The retrieval strategy differs by modality: text sources use hybrid dense-sparse retrieval with cross-encoder reranking, while visual retrieval leverages CLIP embeddings for cross-modal matching.

\medskip
\textbf{(1) Text Retrieval}
\medskip

We employ a hybrid retrieval strategy that combines dense and sparse methods. For dense retrieval, we generate 384-dimensional text embeddings using \texttt{all-MiniLM-L6-v2}, selected for its efficient inference and strong performance on the Massive Text Embedding Benchmark (MTEB)~\citep{wang2020minilm}. Embeddings are indexed with FAISS (Facebook AI Similarity Search) for efficient similarity search. This is complemented by BM25-based sparse retrieval for exact keyword matching. The two retrieval methods are fused using Reciprocal Rank Fusion (RRF) with constant $k=60$. To further refine retrieval quality, we apply a cross-encoder reranker (\texttt{BAAI/bge-reranker-base}), which was chosen for its ability to capture fine-grained query-document relevance and its top-tier performance on the MTEB leaderboard~\citep{xiao2023bge}. This reranker processes the top 20 candidates from the hybrid search.

\medskip
\textbf{(2) Visual Retrieval}
\medskip

Aerial imagery tiles are selected primarily through spatial containment: the system retrieves tiles whose centroids fall within the target ZIP code polygon. If fewer than the required number are found, it falls back to a radius search (5 km, then 30 km) over tiles from the same acquisition period. Optionally, we employ CLIP (ViT-B/32)~\citep{radford2021learningtransferablevisualmodels} for semantic re-ranking, mapping the text query to the visual embedding space to prioritize tiles whose content (e.g., flooded intersections, visible debris) is most relevant to the query.

\medskip
\textbf{(3) Handling Geolocation Uncertainty}
\medskip

A critical challenge in social sensing is the scarcity and unreliability of precise geotags~\citep{mandal_algorithmic_2024}. To address this, our system moves beyond coordinate-based filtering. Instead, we rely on semantic retrieval to identify tweets that are contextually relevant to the target ZIP code, even if they lack explicit GPS metadata. We bypass the constraint of scarce geotags ($<$1\% of tweets) by retrieving posts that mention specific street names, landmarks, or neighborhoods. For example, retrieving reports of ``rescues in Bellaire'' or ``flooding on I-610'' maximizes recall of valid on-the-ground reports for relevant ZIPs even when GPS metadata is missing. We validated retrieval quality by analyzing all retrieved tweets ($N=4{,}140$) across the 207 study queries: 99.7\% contained disaster-relevant keywords, and 60.0\% included explicit geographic mentions (street names or neighborhood references), a result that indicates effective semantic grounding.

\subsection{Split-Pipeline Architecture}
\label{sec:split_pipeline}

A key design decision in our framework is the separation of text and visual reasoning into distinct analysis pipelines (see Figure~\ref{fig:system_architecture}). This split architecture acknowledges that text and imagery capture fundamentally different aspects of disaster impact: text reports describe transient, real-time conditions (rising waters, rescue needs), while imagery captures persistent physical evidence (structural damage, debris). Processing them separately enables modality-specific reasoning before fusion.

\medskip
\textbf{(1) Text Analyst}
\medskip

The Text Analyst module processes retrieved text snippets, including tweets, 311 calls, and peak rainfall observations, and generates a structured analysis, serving as the primary evidence source for local impact. We evaluate four different pre-trained language models in this role: Gemini 2.5 Flash~\citep{gemini2025}, Gemini 3 Flash~\citep{gemini2025}, Qwen 3.5~\citep{qwen2025}, and GPT-5-mini~\citep{openai2025gpt5}.

\medskip
\textbf{(2) Visual Analyst}
\medskip

The Visual Analyst module processes retrieved aerial imagery tiles to produce a comprehensive visual assessment of flood extent and structural damage. Each model serves as its own visual analyst, using its native multimodal capability to ensure consistent visual processing across configurations.

\medskip
\textbf{(3) Temporal-Aware Fusion Engine}
\medskip

We employ a temporal-aware fusion strategy that combines estimates from the Text and Visual Analysts based on when each piece of evidence was captured relative to the disaster timeline. As formally defined below, this module mitigates the ``snapshot vs. peak'' problem by prioritizing text for flood extent (to capture receding waters) while treating visual evidence as additive for structural damage.

\begin{algorithm}[!htbp]
\caption{Asynchronous Multimodal Fusion Logic}
\label{alg:async_fusion}
\begin{algorithmic}
\STATE \textbf{Input:} Text Analysis ($T$), Visual Analysis ($V$) with continuous confidence scores $\text{conf}_T, \text{conf}_V \in [0,1]$;
\STATE \hspace{2.2em} $V$ reports \texttt{conf\_level} $\in \{\text{strong}, \text{partial}, \text{none}, \text{contradicts}, \text{unknown}\}$
\STATE \textbf{Output:} Flood Extent ($E$), Damage Severity ($S$)
\STATE
\STATE // 1. Hazard (Flood Extent): text-primary; visual confirmatory only
\IF{\texttt{conf\_level} $\in \{\text{contradicts},\, \text{none}\}$}
    \STATE $E \leftarrow T_{\mathrm{extent}}$ \COMMENT{Water receded post-event; discard visual}
\ELSIF{\texttt{conf\_level} $= \text{strong}$}
    \STATE $w_T \leftarrow 0.75 + (\text{conf}_T - 0.5)\cdot 0.1;\quad w_V \leftarrow 0.25 + (\text{conf}_V - 0.5)\cdot 0.1$
    \STATE $E \leftarrow (w_T \cdot T_{\mathrm{extent}} + w_V \cdot V_{\mathrm{extent}})\;/\;(w_T + w_V)$ \COMMENT{$w_T \in [0.70,0.80]$}
\ELSIF{\texttt{conf\_level} $= \text{partial}$}
    \STATE $w_T \leftarrow 0.80 + (\text{conf}_T - 0.5)\cdot 0.1;\quad w_V \leftarrow 0.20 + (\text{conf}_V - 0.5)\cdot 0.1$
    \STATE $E \leftarrow (w_T \cdot T_{\mathrm{extent}} + w_V \cdot V_{\mathrm{extent}})\;/\;(w_T + w_V)$ \COMMENT{$w_T \in [0.75,0.85]$}
\ELSE
    \STATE $E \leftarrow T_{\mathrm{extent}}$ if $T_{\mathrm{extent}}>0$, else $V_{\mathrm{extent}} \times \min(0.8,\; \text{conf}_V + 0.2)$ \COMMENT{visual fallback when text absent}
\ENDIF
\STATE
\STATE // 2. Consequence (Damage): text is floor; visual is additive only
\IF{$\text{conf}_V \leq 0.5$}
    \STATE $S_{\mathrm{fused}} \leftarrow T_{\mathrm{damage}}$ \COMMENT{Low visual confidence: ignore visual}
\ELSE
    \STATE $S_{\mathrm{fused}} \leftarrow \text{WeightedAvg}(T_{\mathrm{damage}}, V_{\mathrm{damage}};\; \text{conf}_T, \text{conf}_V)$
\ENDIF
\STATE $S \leftarrow \max(S_{\mathrm{fused}},\; T_{\mathrm{damage}})$ \COMMENT{Visual can boost but never reduce text estimate}
\RETURN $E, S$
\end{algorithmic}
\end{algorithm}

To formally define the integration of asynchronous modalities, we employ a temporal-aware fusion logic that respects when each data stream was collected:

As shown in Algorithm~\ref{alg:async_fusion}, this logic ensures that a clear post-event image does not veto a high-confidence text report of earlier flooding (preventing false negatives in extent), while ensuring that visible structural debris in imagery contributes to the damage score even if text reports are sparse (preventing false negatives in damage). The base weights $(b_T, b_V)$ are $(0.75, 0.25)$ for strong confirmation and $(0.80, 0.20)$ for partial, with both adjusted by the respective analyst confidence scores ($\text{conf}_T$, $\text{conf}_V \in [0,1]$ extracted from each analyst's structured JSON output). The key structural constraint is $b_T > b_V$ in all branches: text always outweighs visual evidence, regardless of confidence.

\subsection{Reasoning and Alignment Strategy}
\label{sec:reasoning_strategy}

Beyond the architectural fusion rules in Algorithm~\ref{alg:async_fusion}, we embed domain knowledge directly into the model prompts through two strategies:

\textbf{Temporal Context:} Text sources typically report conditions during peak flooding (e.g., August 27--28), while aerial imagery was captured post-event (August 31, after most floodwaters had receded). The system prompt explicitly informs the model of this temporal gap, instructing it that a visually dry image does not mean flooding did not occur. The model receives peak rainfall accumulation (inches) via spatial priors as a persistent, objective flood indicator. Consequently, the system prioritizes concurrent text reports for flood extent assessment.
    
\textbf{Metric-Aligned Reasoning:} A critical failure mode in standard RAG is metric misalignment, where LLMs interpret ``damage severity'' as a qualitative intensity score (e.g., ``catastrophic'' = 90\%) rather than the specific statistical definition of the ground truth. We introduce a Metric Alignment Constraint in the system prompt that explicitly defines \texttt{damage\_severity\_pct} as the average damage ratio across all structures in the ZIP code (0--100\%). This forces the model to perform an implicit denominator estimation (e.g., ``reports show 50 flooded homes in a neighborhood of around 500, implying about 10\% severity'') rather than reacting solely to the emotional intensity of individual reports. This constraint anchors the model's output to the ground-truth metric definition, ensuring that domain-specific statistical grounding complements retrieval quality.

%==============================================================================
% EXPERIMENTS
%==============================================================================
\section{Experiments and Results}
\label{sec:experiments}

\subsection{Dataset}
\label{sec:dataset}

We evaluate CrisiSense-RAG on data from Hurricane Harvey, which made landfall on August 25, 2017, causing unprecedented flooding in the Greater Houston area. Harvey stalled over the region for four days, dropping over 60 inches of rain in some areas and affecting over 300,000 structures. Table~\ref{tab:dataset_stats} summarizes our data sources.

% Dataset Statistics Table
\begin{table}[!htbp]
    \centering
    \caption{Dataset statistics for the Hurricane Harvey evaluation.}
    \label{tab:dataset_stats}
    \begin{tabular}{lrr}
        \toprule
        \textbf{Data Source}            & \textbf{Count} & \textbf{Coverage} \\
        \midrule
        Aerial Imagery Tiles            & 3,507          & Greater Houston   \\
        Image Captions      & 3,507          & Greater Houston   \\
        Tweets (filtered)               & 458,453        & Harris County     \\
        311 Emergency Calls             & 26,107         & City of Houston   \\
        Precipitation Sensors           & 8,397 daily records & Harris County     \\
        Property Damage Extent (ground truth)    & 72,755         & 139 ZIPs          \\
        \midrule
        Queries (core, with imagery) & 110 & Harris County \\
        Queries (full study area) & 207 & Greater Houston \\
        Flood Depth Grid (ground truth) & ZIP-level flooded area for 207 queries & Harvey footprint  \\
        \bottomrule
    \end{tabular}
\end{table}

We employ a dual ground-truth framework to evaluate the distinct Hazard and Consequence outputs of CrisiSense-RAG:

\textbf{Hazard Ground Truth (Flood Extent)}: We use the FEMA Harvey Flood Depth Grid~\citep{fema_harvey_flood_depths_grid_2023} to compute the percentage of each ZIP code covered by water (\texttt{flooded\_pct}). This is the target for the model's \texttt{flood\_extent\_pct} prediction.

\textbf{Damage Ground Truth (Structural Severity)}: We utilize the Property Damage Extent (PDE) metric, proposed and computed by \citet{ma2024non}, for 139 ZIP codes to quantify structural impact. The ground truth is \texttt{mean\_pde} $\times$ 100, where \texttt{mean\_pde} represents the average damage per building across all buildings in a ZIP code (0--1 range). This is the target for the model's \texttt{damage\_severity\_pct} prediction, which we explicitly define as ``average damage per building'' to ensure semantic alignment. PDE coverage does not span the full 207 ZIPs, so 68 ZIPs lack damage ground truth and are treated as missing in the spatial maps.

We construct evaluation queries for 207 unique ZIP codes in the Greater Houston region, all targeting the primary Hurricane Harvey impact window (August 25--September 1, 2017). Of these, 110 ZIP codes have aerial imagery tile coverage and form the core evaluation area used for all experiments: these ZIPs support all four modality configurations (Text-Only, Text+Caption, No-Tweets, Full Multimodal), and the main ablation results are reported on this set. The remaining 97 ZIP codes have no indexed imagery tiles, reducing all configurations to text-only inference for those areas. We report text-only performance on these peripheral ZIPs and on the combined 207-ZIP study area as a supplementary analysis in Section~\ref{sec:suppl_207}, demonstrating broader geographic generalization of the text retrieval pipeline.

\subsection{Ablation Study Design}
\label{sec:baselines}

We conduct a systematic ablation study across four configurations to isolate the contribution of each modality. As detailed in Section~\ref{sec:dataset}, we evaluate on the 110 ZIP codes with imagery coverage, ensuring a fair comparison where all four configurations operate on the same queries; text-only results for the remaining 97 peripheral ZIPs are reported in Section~\ref{sec:suppl_207}.

\textbf{Full Multimodal RAG:} Complete split-pipeline architecture with separate Text Analyst and Visual Analyst modules. To avoid double-counting the same imagery signal, pre-generated captions are excluded from the text stage in this configuration; the visual branch sees the raw imagery directly. Each model serves as its own visual analyst, using its native multimodal capability.

\textbf{No Social Media (No-Tweets):} Full multimodal configuration with tweets removed from retrieval, isolating the contribution of social media relative to 311 calls, rainfall data, and aerial imagery alone. Pre-generated captions are also excluded (same as the multimodal configuration) to maintain a clean ablation.

\textbf{Text+Caption RAG:} Text sources augmented by pre-generated image captions. Captions are generated from aerial imagery using Gemini 2.0 Flash and indexed as searchable text. This tests whether semantic descriptions of imagery can bridge the modality gap without direct visual analysis.

\textbf{Text-Only RAG:} Baseline using only text sources (tweets, 311 calls, and peak rainfall data) without any imagery input.

\subsection{Evaluation Metrics}
\label{sec:metrics}

We evaluate prediction accuracy using Mean Absolute Error (MAE) for both Hazard (Flood Extent) and Consequence (Damage Severity). Ground truth for each metric is defined as follows.

For each ZIP code $z_i$, the ground truth flood extent is the percentage of the ZIP area inundated according to the FEMA flood depth raster:
\begin{equation}
    y_i^{\mathrm{ext}} = \frac{|A(z_i) \cap F|}{|A(z_i)|} \times 100
    \label{eq:gt_extent}
\end{equation}
where $A(z_i)$ is the total area of ZIP $z_i$ and $F$ is the FEMA flood inundation footprint.

The ground truth damage severity is the mean Point Damage Estimate (PDE) across all buildings within the ZIP:
\begin{equation}
    y_i^{\mathrm{dmg}} = \frac{1}{|B_i|} \sum_{b \in B_i} \mathrm{PDE}_b \times 100
    \label{eq:gt_damage}
\end{equation}
where $B_i$ is the set of buildings in $z_i$ and $\mathrm{PDE}_b \in [0, 1]$ is the fractional structural damage estimate for building $b$.

MAE is then computed as the average absolute deviation between model predictions $\hat{y}_i$ and ground truth $y_i$ (both in percentage points, 0--100):
\begin{equation}
    \mathrm{MAE} = \frac{1}{N} \sum_{i=1}^{N} \left|\hat{y}_i - y_i\right|
    \label{eq:mae}
\end{equation}
We report separate MAE scores for flood extent (Eq.~\ref{eq:gt_extent}) and damage severity (Eq.~\ref{eq:gt_damage}), allowing us to assess performance on each distinct output of our dual-metric system.

\subsection{Main Results}
\label{sec:main_results}

Table~\ref{tab:main_results} reports full multimodal RAG performance across four model backends on the 110 ZIP codes with imagery coverage ($N=110$). All models operate in a zero-shot setting and process both text context and aerial imagery using their native multimodal capability. A per-configuration breakdown comparing all four ablation conditions is provided in Table~\ref{tab:ablation_configs} and Figure~\ref{fig:results_barchart} in Section~\ref{sec:ablation_study}; text-only results on the full 207-ZIP study area are in Section~\ref{sec:suppl_207}.

% Main Results Table — Multimodal only
\begin{table}[!htbp]
    \centering
    \caption{Full Multimodal RAG results on the 110 ZIPs with imagery coverage. Flood Extent MAE: $N=110$; Damage MAE: $N=104$ (ZIPs with PDE ground truth only). Bootstrapped 95\% CI. Lower is better.}
    \label{tab:main_results}
    \begin{tabular}{lcc}
        \toprule
        \textbf{Model} & \textbf{Extent MAE (95\% CI)} & \textbf{Damage MAE (95\% CI)} \\
        \midrule
        Gemini 2.5 Flash & 12.98 (11.20--14.73) & 12.31 (10.38--14.33) \\
        Gemini 3 Flash   & 11.11 (9.53--12.75)  & \textbf{10.14} (8.69--11.61)  \\
        Qwen 3.5    & 19.64 (17.29--21.93) & 11.23 (9.81--12.77) \\
        GPT-5-mini       & \textbf{8.86} (7.47--10.26)  & 14.47 (12.51--16.43) \\
        \bottomrule
    \end{tabular}
\end{table}

\begin{figure}[p]
    \centering
    % Row 1 — Ground Truth
    \begin{subfigure}[b]{0.49\textwidth}
        \centering
        \includegraphics[width=\textwidth]{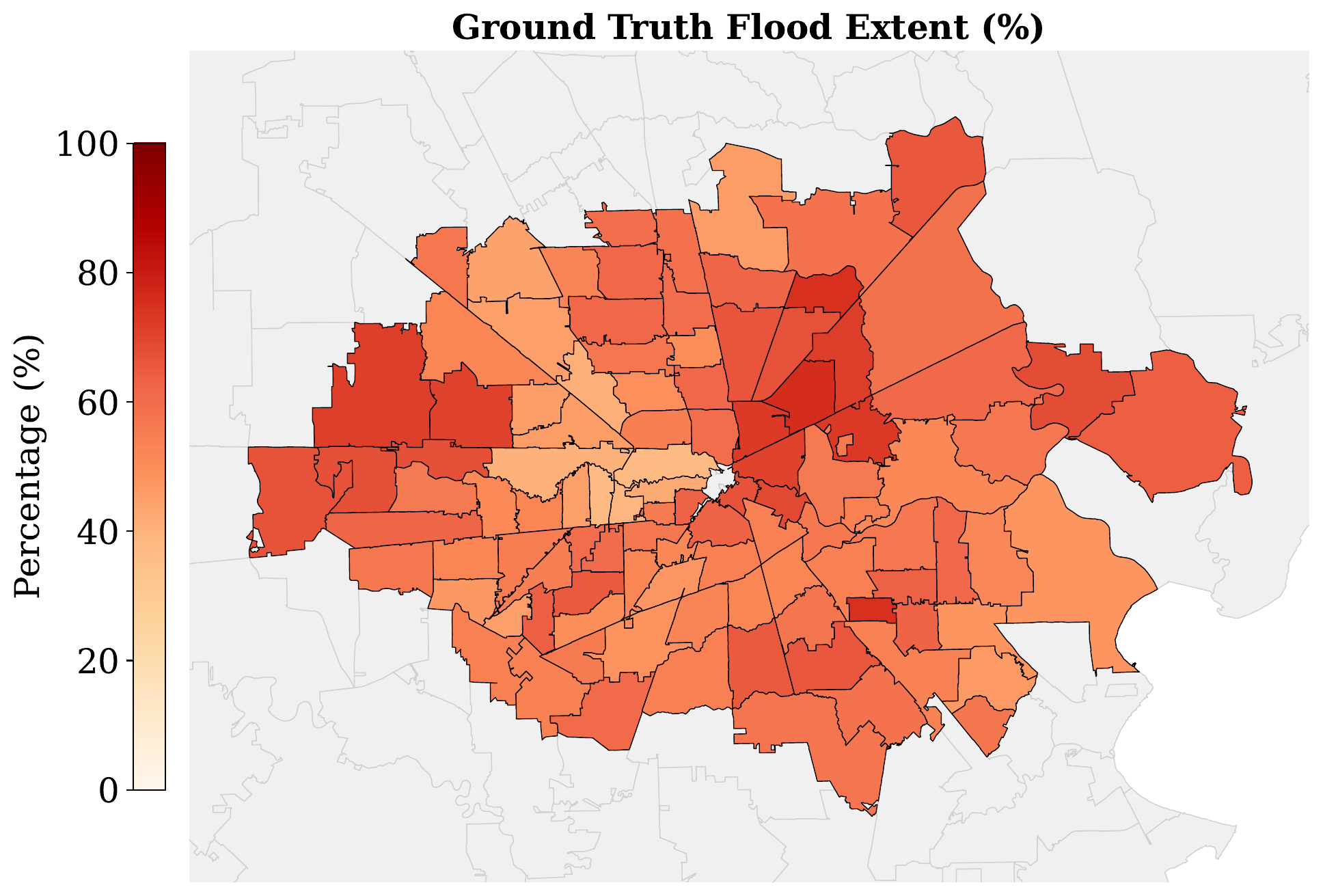}
        \caption{GT Flood Extent}
        \label{fig:map_extent_gt}
    \end{subfigure}
    \hfill
    \begin{subfigure}[b]{0.49\textwidth}
        \centering
        \includegraphics[width=\textwidth]{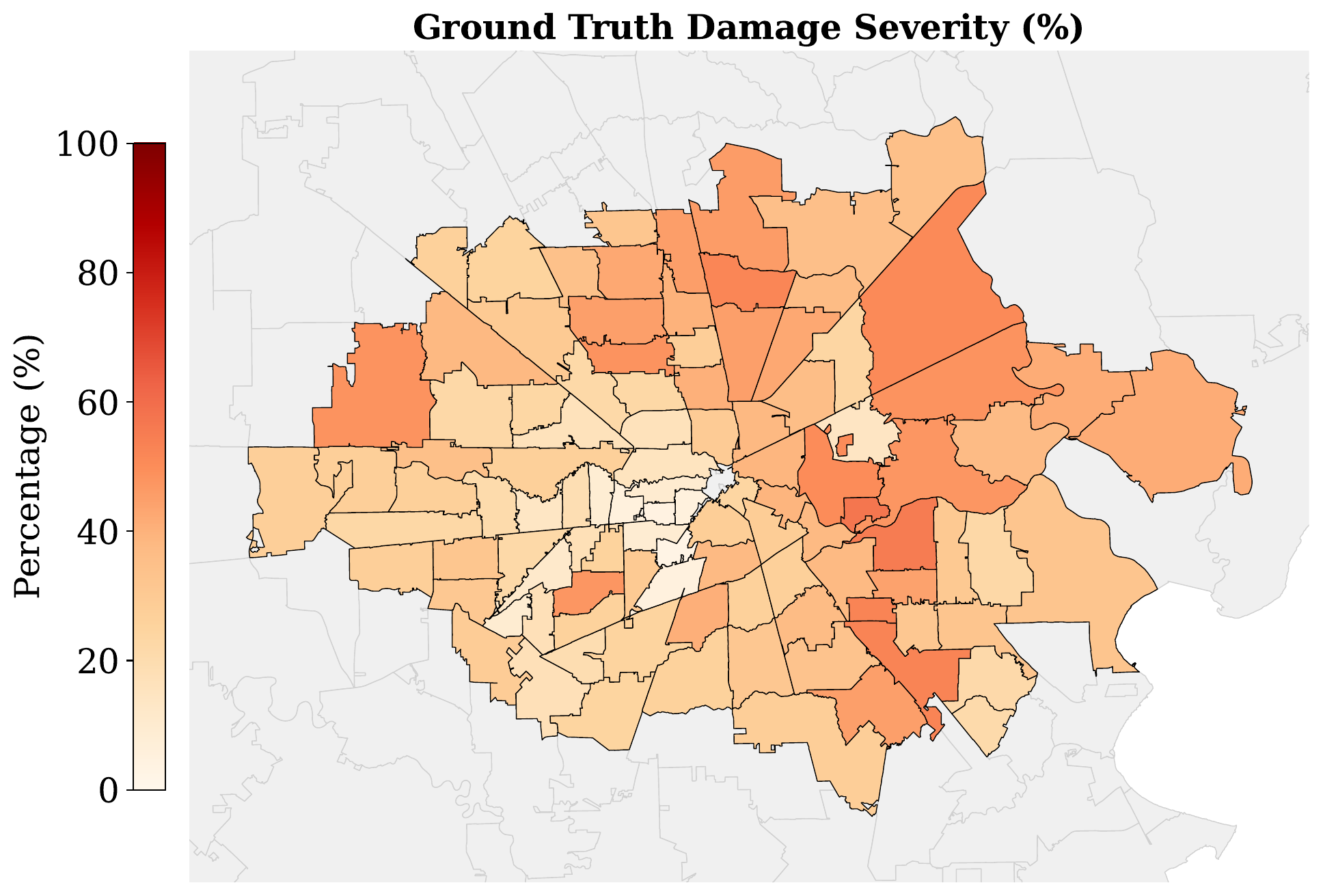}
        \caption{GT Damage Severity}
        \label{fig:map_damage_gt}
    \end{subfigure}

    % \vspace{0.3em}

    % Row 2 — Predictions
    \begin{subfigure}[b]{0.49\textwidth}
        \centering
        \includegraphics[width=\textwidth]{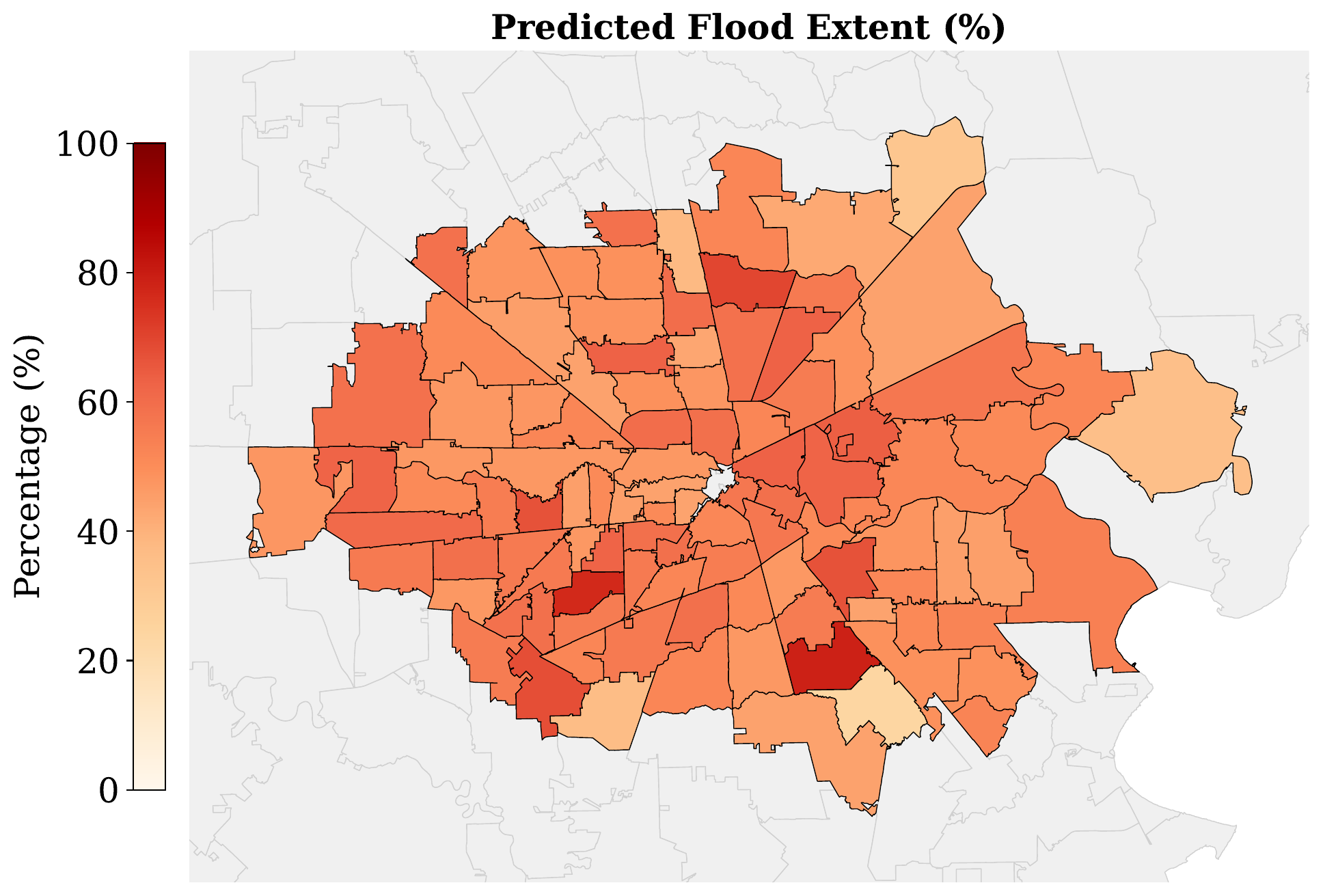}
        \caption{Predicted Flood Extent (GPT-5-mini)}
        \label{fig:map_extent_pred}
    \end{subfigure}
    \hfill
    \begin{subfigure}[b]{0.49\textwidth}
        \centering
        \includegraphics[width=\textwidth]{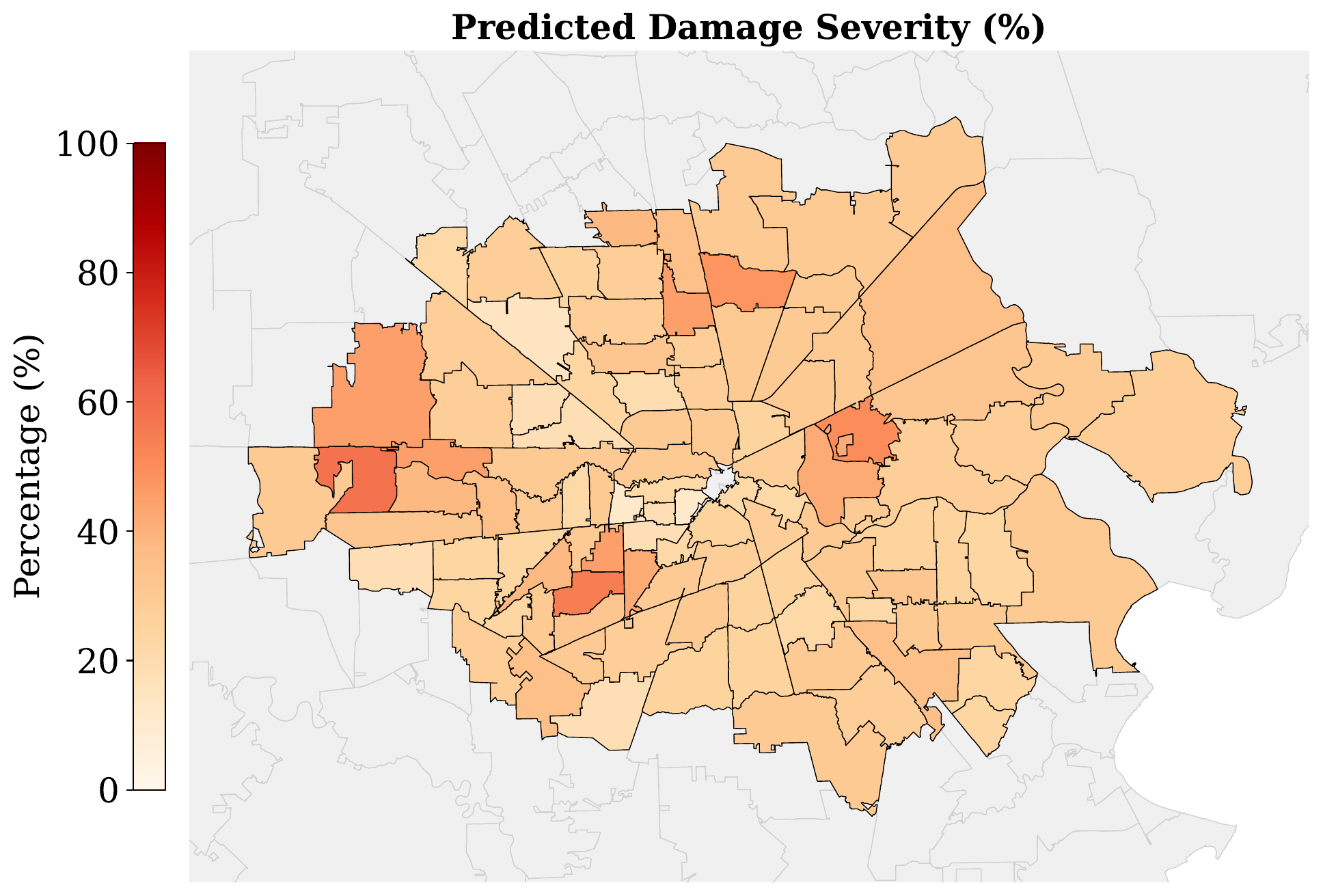}
        \caption{Predicted Damage Severity (Gemini 3 Flash)}
        \label{fig:map_damage_pred}
    \end{subfigure}

    % \vspace{0.3em}

    % Row 3 — Residuals
    \begin{subfigure}[b]{0.49\textwidth}
        \centering
        \includegraphics[width=\textwidth]{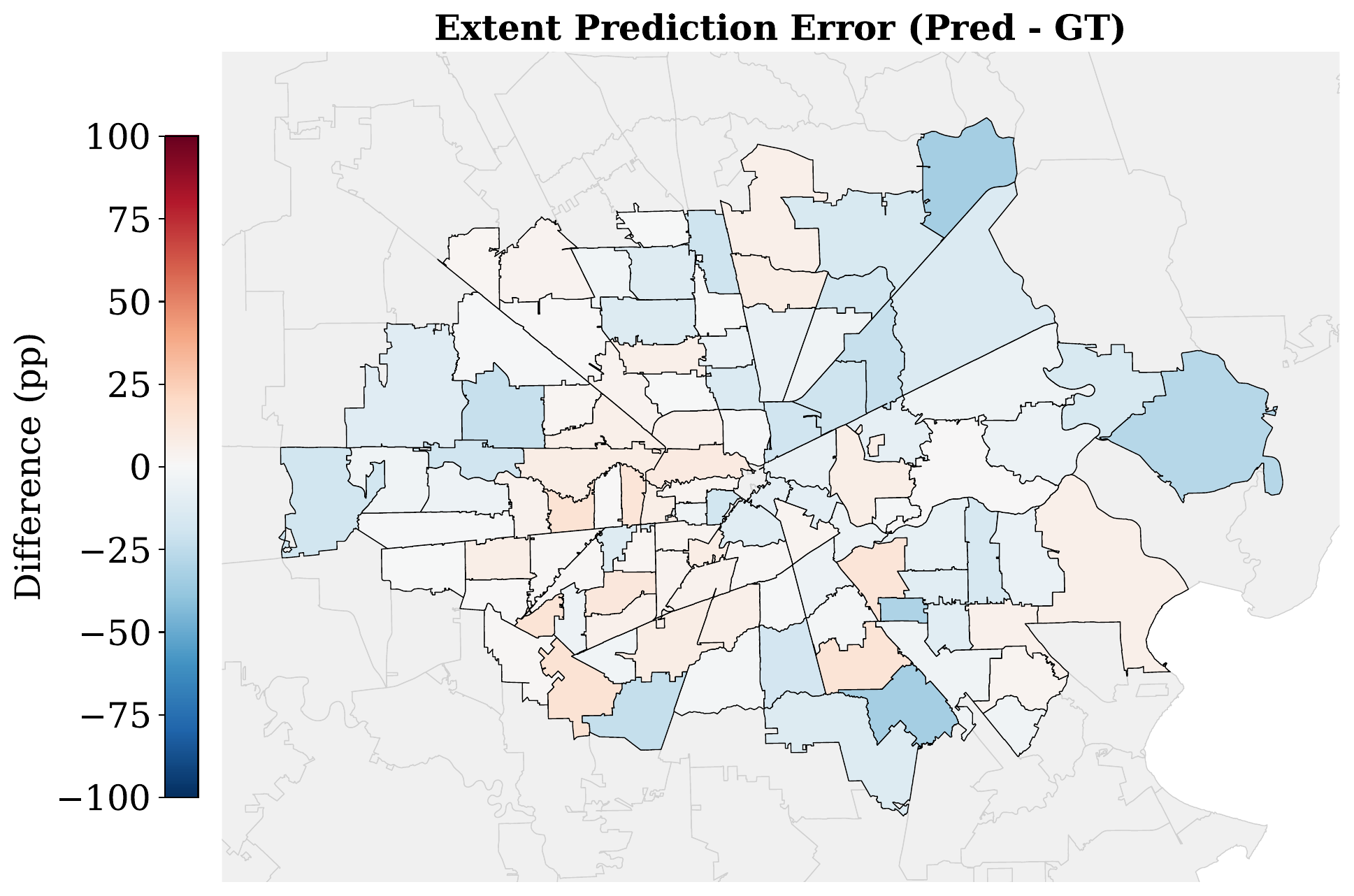}
        \caption{Residual: Flood Extent (Pred $-$ GT)}
        \label{fig:map_extent_error}
    \end{subfigure}
    \hfill
    \begin{subfigure}[b]{0.49\textwidth}
        \centering
        \includegraphics[width=\textwidth]{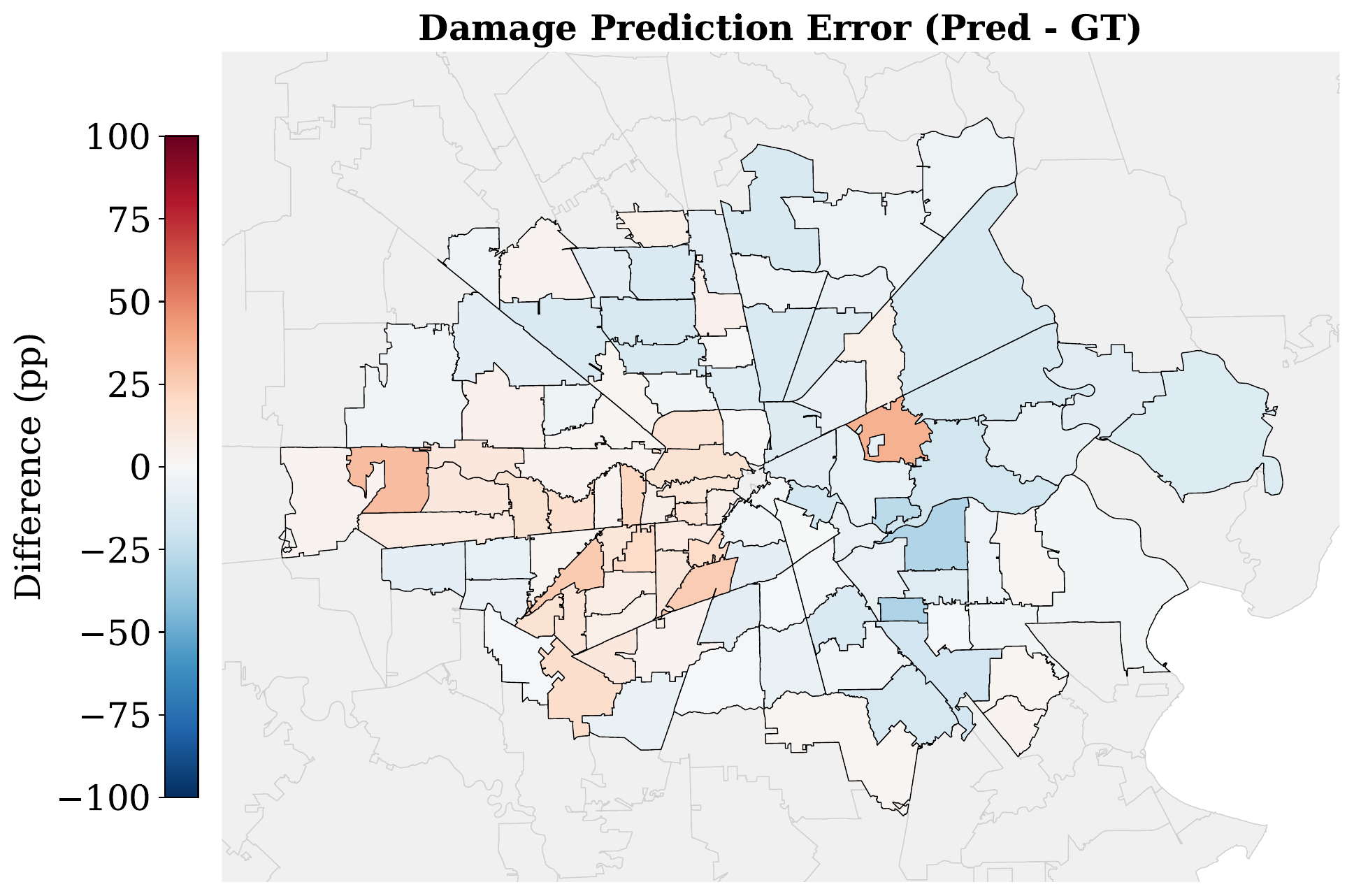}
        \caption{Residual: Damage Severity (Pred $-$ GT)}
        \label{fig:map_damage_error}
    \end{subfigure}

    \caption{Ground truth (top), predictions (middle), and residuals (bottom) for Flood Extent (left, GPT-5-mini Multimodal) and Damage Severity (right, Gemini 3 Flash Multimodal) across 104 ZIP codes. Blue indicates underestimation and red overestimation.}
    \label{fig:spatial_maps}
\end{figure}

All four model backends achieve operationally useful zero-shot performance, with Extent MAE ranging from 8.86\% (GPT-5-mini) to 19.64\% (Qwen 3.5) and Damage MAE ranging from 10.14\% (Gemini 3 Flash) to 14.47\% (GPT-5-mini). Figure~\ref{fig:spatial_maps} shows the spatial distribution of predictions and residuals for the best-performing multimodal configuration per metric: flood extent residuals show slight systematic underestimation in peripheral ZIP codes where social media evidence is sparse, and damage residuals show localized overestimation near dense urban areas. A detailed breakdown of what each modality contributes is provided in Section~\ref{sec:ablation_study}.

To validate generalization across model families, we evaluated the split-pipeline design with four distinct text backends: Gemini 2.5 Flash and Gemini 3 Flash (closed-source), Qwen 3.5, and GPT-5-mini. While performance varies by model and configuration, the full multimodal results show that all four backends can produce quantitative zero-shot predictions without any disaster-specific training.

\subsection{Spatial Distribution of Impact}
\label{sec:spatial_results}

Figure~\ref{fig:spatial_maps} illustrates the spatial alignment between ground truth and model predictions for both flood extent and damage severity across the core 104 ZIP codes with PDE coverage. Visually, the model effectively captures the broad geographic distribution of flooding, accurately identifying high-impact areas along major waterways and reservoirs. For damage severity, the predictions show strong spatial agreement with the clusters of structural damage recorded in the Property Damage Extent (PDE). Notably, by focusing the visualization and evaluation on the core area where ground truth coverage is most complete, we observe that the system successfully disambiguates localized hazard reports into a coherent geographic assessment. While the model occasionally over-predicts or under-predicts impact in areas with dense social media activity, the overall spatial correspondence validates our approach's utility for rapid situational awareness.

\textbf{Spatial diagnostics.} To quantify the geographic structure of prediction errors visible in the residual maps (Figure~\ref{fig:spatial_maps}, bottom row), we computed Moran's~$I$ statistic on the signed residuals using Queen contiguity weights (999 permutations). Flood extent residuals exhibit moderate positive spatial autocorrelation ($I = 0.175$, $p = 0.003$), indicating that underestimation errors cluster geographically and predominantly in peripheral ZIP codes at the outer edge of the Harvey impact zone, where social media coverage is sparse. Damage severity residuals show stronger clustering ($I = 0.391$, $p = 0.001$), consistent with the spatial concentration of structural damage along specific flood pathways and the tendency of tweet-based evidence to propagate across neighboring ZIP codes when reports lack precise location anchors. Both statistics are substantially above the expected value under spatial randomness ($E[I] \approx -0.009$), confirming that the error structure is not random noise but reflects systematic geographic patterns tied to data availability and social media density.

%==============================================================================
\subsection{Comparison with Related Approaches}
\label{sec:cross_study_comparison}

\begin{table}[!htbp]
    \centering
    \caption{Cross-study comparison of damage estimation methods. All studies target Hurricane Harvey except Arachchige \& Pradhan (Florida hurricanes). ``Approx.\ MAE'' denotes normalized error estimates converted from classification accuracy; see text for details. For CrisiSense-RAG, we report the best observed damage MAE across evaluated configurations to provide an optimistic reference point. Comparisons remain approximate due to differences in resolution, supervision, and task formulation.}\label{tab:cross_study}
    \begin{tabularx}{\textwidth}{L l l r L}
        \toprule
        \textbf{Method} & \textbf{Resolution} & \textbf{Supervision} & \textbf{Approx.\ MAE} & \textbf{Data Sources} \\
        \midrule
        R2RAG-Flood \cite{huangR2RAGFloodReasoningreinforcedTrainingfree2026} & HUC12 & Zero-shot & $17.5$--$22.5\%$ & Tabular predictors, LLM reasoning \\
        Arachchige \& Pradhan \cite{maha_arachchige_ai_2025} & ZCTA & Supervised & $11.3\%$ & Insurance claims (1985--2024) \\
        FloodDamageCast \cite{liu_flooddamagecast_2024} & 500\,m grid & Supervised & ${\sim}14.3\%$ & Built env., topo., hydro.\ features \\
        Flood-DamageSense \cite{ho_multimodal_2025} & Building & Supervised & --- & SAR/InSAR + optical imagery \\
        \textbf{CrisiSense-RAG (ours)} & ZIP code & Zero-shot & $10.09\%$ & Social media, 311 calls, rainfall, aerial imagery \\

        \bottomrule
    \end{tabularx}
\end{table}

To further contextualize our results, we compare against four recent approaches that address flood damage estimation in overlapping study areas and with partially comparable metrics. We emphasize that these are not direct apples-to-apples comparisons: the studies differ in spatial resolution (ZIP-code vs.\ grid or building level), supervision paradigm (zero-shot vs.\ supervised), and task formulation (continuous regression vs.\ ordinal classification). Nonetheless, the comparison illuminates where our zero-shot framework stands relative to systems that benefit from task-specific training or higher-resolution inputs.

Where prior work reports $K$-level ordinal classification accuracy rather than a continuous MAE, we approximate a normalized MAE as:
\begin{equation}
    \text{Approx.\ MAE} \approx (1 - \text{accuracy}) \times \bar{d} \times 100\%
    \label{eq:ordinal_mae}
\end{equation}
where $\bar{d}$ is the expected ordinal displacement of a misclassification, normalized by the maximum possible displacement~\cite{baccianella2009evaluation,cardoso2007learning}. For three-level PDE categories with equal class spacing (0\%, 50\%, 100\%), adjacent errors contribute a displacement of 0.5 and non-adjacent errors contribute 1.0, giving a theoretical range of $\bar{d} \in [0.5, 1.0]$. Because the ratio of adjacent to non-adjacent errors is not reported, we narrow this range by conservatively assuming the majority (80--90\%) of errors are between adjacent classes (the assumption that maximizes $\bar{d}$ and thus yields the largest, most pessimistic MAE bound), yielding $\bar{d} \in [0.55, 0.60]$ and the MAE ranges reported in Table~\ref{tab:cross_study}. These estimates are approximate and are intended only for broad ordering comparisons.

Some prior works are based on zero-shot RAG frameworks. For example, R2RAG-Flood~\cite{huangR2RAGFloodReasoningreinforcedTrainingfree2026} is a reasoning-reinforced, training-free RAG framework that predicts Property Damage Extent (PDE) at the HUC12 watershed scale in Harris County. Operating across seven LLM backbones, it attains overall accuracy of $0.613$--$0.668$ on three-level PDE categories. Converting their classification error to a normalized MAE on the same ordinal scale yields approximately $17.5\%$--$22.5\%$. Our best observed damage MAE of $10.09\%$ (Gemini 3 Flash, Text-Only) outperforms this reasoning-based system, despite operating at a coarser ZIP-code resolution and relying on heterogeneous real-time data streams rather than curated tabular predictors.

A number of supervised learning approaches have also been proposed recently. 
FloodDamageCast~\cite{liu_flooddamagecast_2024} employs a supervised LightGBM model augmented with CTGAN-based data augmentation, using the same PDE labels at a $500$m $\times$ $500$m grid resolution in Harris County. It achieves an overall accuracy of $0.714$ and damage class accuracy of $0.859$. Their classification error rate of $0.286$ on three ordinal levels corresponds to a normalized MAE of approximately $14.3\%$. Arachchige and Pradhan~\cite{maha_arachchige_ai_2025} develop a fully supervised stacked ensemble model trained on nearly four decades of building-level insurance claims data (1985--2024) to predict hurricane flood damage at the ZIP Code Tabulation Area (ZCTA) level, achieving an MAE of $11.3\%$. This represents a strong supervised reference point for area-level damage prediction. Our best observed zero-shot result ($10.09\%$) is competitive with this supervised benchmark despite requiring no task-specific training data.

Frameworks with finer resolution (e.g., building level) are also very well explored. Flood-DamageSense~\cite{ho_multimodal_2025} is a supervised deep-learning framework using a multimodal Mamba backbone that fuses SAR/InSAR with optical imagery for building-level damage classification in Harris County. It reports a mean F1 improvement of $19$ percentage points over prior baselines. Direct numerical comparison with our system is not meaningful due to the fundamentally different granularity (individual buildings vs.\ ZIP codes) and metric (F1 vs.\ MAE). However, this work highlights the complementary nature of building-level and area-level assessment. While Flood-DamageSense provides high-precision structural localization under cloud cover, CrisiSense-RAG provides rapid ``macro'' situational awareness. It achieves this by integrating social media, emergency calls, and rainfall data to capture community-level impacts often invisible to remote sensing alone.

Taken together, these comparisons show that CrisiSense-RAG's zero-shot damage estimates (best observed MAE = 10.09\%) are competitive with purpose-built supervised systems and outperform existing zero-shot approaches, despite the absence of training data, coarser spatial resolution, and the operational constraint of real-time deployment. Our framework approaches the supervised reference point set by~\cite{maha_arachchige_ai_2025} for area-level damage prediction, while the zero-shot RAG comparison~\cite{huangR2RAGFloodReasoningreinforcedTrainingfree2026} confirms that multimodal fusion of heterogeneous data sources outperforms reasoning over curated tabular features alone.

\section{Analysis}\label{sec:analysis}

\subsection{Ablation Study}\label{sec:ablation_study}

\begin{figure}[!htbp]
    \centering
    \includegraphics[width=\textwidth]{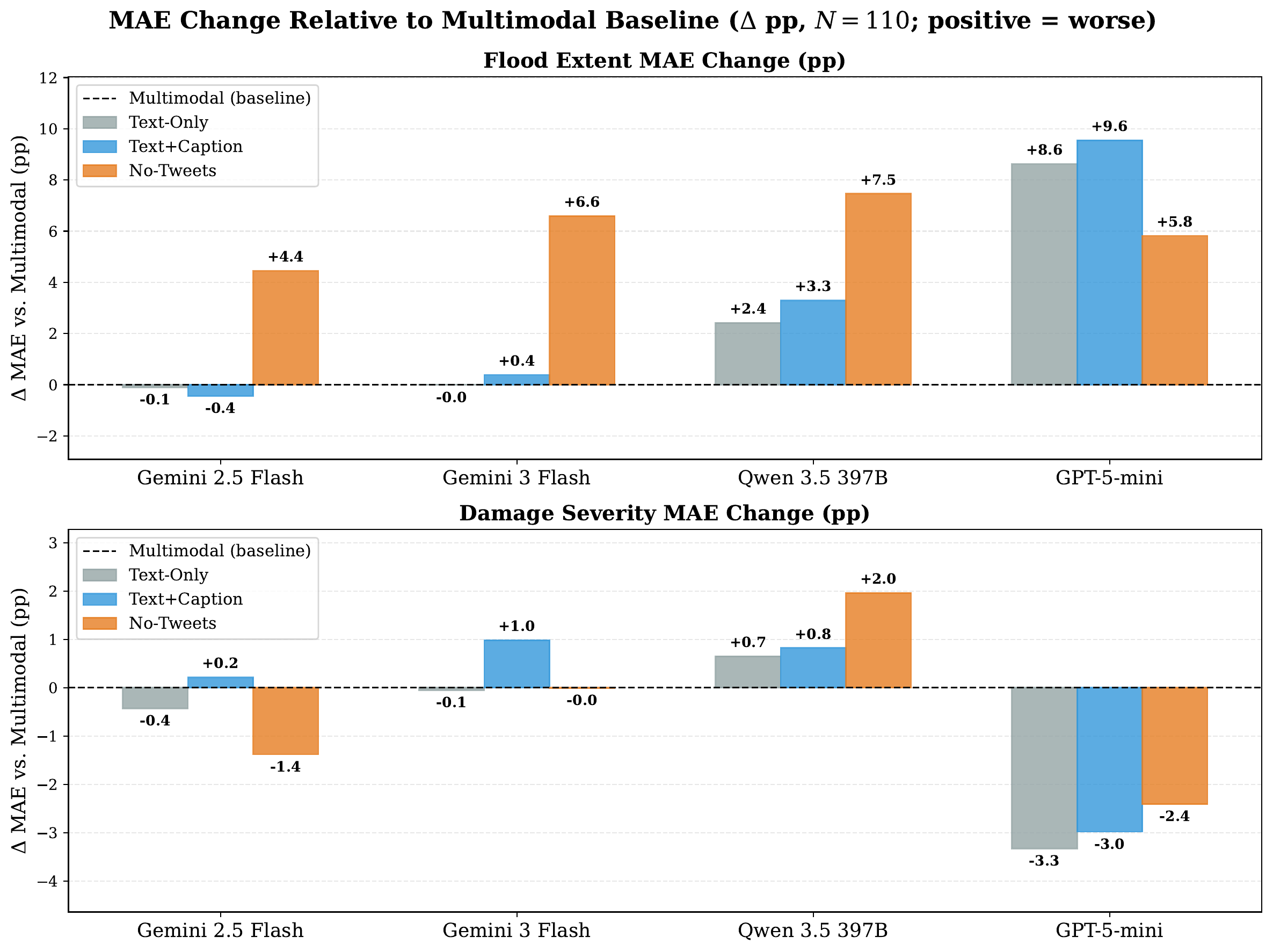}
    \caption{MAE change relative to the Full Multimodal baseline for Text-Only, Text+Caption, and No-Tweets configurations across four model backends. Absolute MAE values are reported in Table~\ref{tab:ablation_configs}.}
    \label{fig:results_barchart}
\end{figure}

We analyze how performance changes as individual data sources are removed or substituted in the full pipeline, isolating the contribution of aerial imagery, pre-generated captions, and social media (configurations defined in Section~\ref{sec:baselines}). Table~\ref{tab:ablation_configs} reports MAE for all four configurations; Figure~\ref{fig:results_barchart} visualizes the MAE change relative to the Full Multimodal baseline.

\begin{table}[!htbp]
    \centering
    \caption{Ablation results across four configurations on the 110 ZIPs with imagery coverage. Flood Extent MAE: $N=110$; Damage MAE: $N=104$ (ZIPs with PDE ground truth only). Bootstrapped 95\% CI. Bold = best per model per metric.}\label{tab:ablation_configs}
    \begin{tabular}{lcc}
        \toprule
        \textbf{Model \& Configuration} & \textbf{Extent MAE (95\% CI)} & \textbf{Damage MAE (95\% CI)} \\
        \midrule
        \multicolumn{3}{l}{\textit{Gemini 2.5 Flash}} \\
        \quad Text-Only   & 12.88 (11.29--14.38) & 11.88 (10.07--13.90) \\
        \quad Text+Caption & \textbf{12.54} (10.82--14.31) & 12.52 (10.61--14.56) \\
        \quad No-Tweets   & 17.42 (15.29--19.79) & \textbf{10.93} (9.43--12.48) \\
        \quad Multimodal  & 12.98 (11.20--14.73) & 12.31 (10.38--14.33) \\

        \midrule
        \multicolumn{3}{l}{\textit{Gemini 3 Flash}} \\
        \quad Text-Only   & \textbf{11.10} (9.62--12.65)  & \textbf{10.09} (8.62--11.56) \\
        \quad Text+Caption & 11.50 (9.92--13.25) & 11.12 (9.63--12.62) \\
        \quad No-Tweets   & 17.70 (15.83--19.67) & 10.13 (8.92--11.45) \\
        \quad Multimodal  & 11.11 (9.53--12.75)  & 10.14 (8.69--11.61) \\

        \midrule
        \multicolumn{3}{l}{\textit{Qwen 3.5}} \\
        \quad Text-Only   & 22.06 (19.92--24.11) & 11.88 (10.25--13.59) \\
        \quad Text+Caption & 22.93 (20.85--24.96) & 12.05 (10.30--13.89) \\
        \quad No-Tweets   & 27.10 (24.81--29.30) & 13.19 (11.61--14.89) \\
        \quad Multimodal  & \textbf{19.64} (17.29--21.93) & \textbf{11.23} (9.81--12.77) \\

        \midrule
        \multicolumn{3}{l}{\textit{GPT-5-mini}} \\
        \quad Text-Only   & 17.48 (15.79--19.16) & \textbf{11.14} (9.57--12.76) \\
        \quad Text+Caption & 18.41 (16.54--20.31) & 11.49 (9.90--13.21) \\
        \quad No-Tweets   & 14.68 (13.07--16.36) & 12.06 (10.41--13.85) \\
        \quad Multimodal  & \textbf{8.86} (7.47--10.26)  & 14.47 (12.51--16.43) \\

        \bottomrule
    \end{tabular}
\end{table}

\medskip
\textbf{(1) Aerial imagery} 
\medskip

Full multimodal RAG reduces flood extent MAE for two out of four models: GPT-5-mini improves from 17.48\% to 8.86\% ($-$8.62 pp) and Qwen 3.5 from 22.06\% to 19.64\% ($-$2.42 pp). Both Gemini variants show negligible change (Gemini 2.5 Flash: $+$0.10 pp; Gemini 3 Flash: $+$0.01 pp), indicating that their text-only predictions are already well-calibrated for this metric. This benefit reflects a natural fit: inundation manifests as large, spatially coherent water bodies that are visually unambiguous in overhead tiles, making imagery a direct confirmation or correction of text-reported flood extent. Structural damage requires a fundamentally different evidence type (building-level changes such as roof deformations, debris accumulation, or facade discoloration) that is difficult to resolve at the available tile resolution. The post-event acquisition timing compounds this: tiles captured three to four days after peak flooding depict receded water rather than the damage state at peak impact. Damage estimation therefore relies primarily on 311 emergency calls, which directly log building-level flooding complaints, and on FEMA historical knowledge encoding damage patterns from prior events, both of which have no visual counterpart in the available imagery. Accordingly, three out of four models show no improvement or degradation in damage MAE when adding imagery (Gemini 2.5 Flash: $+$0.43 pp; Gemini 3 Flash: $+$0.07 pp; GPT-5-mini: $+$3.33 pp), with only Qwen showing a marginal gain ($-$0.65 pp). For GPT-5-mini specifically, the ambiguous post-recession tiles introduce visual noise that its fusion behavior amplifies, yielding the largest degradation; Gemini and Qwen are more robust to this noise.

\medskip
\textbf{(2) Caption substitution (Text+Caption vs.\ Multimodal).} 
\medskip

The Text+Caption configuration replaces direct visual analysis with pre-generated text descriptions of the same aerial tiles, testing whether captions can serve as a low-cost substitute for the full visual branch. The answer is model-dependent. For flood extent, Gemini variants show negligible degradation relative to Multimodal ($-$0.44 pp for Gemini 2.5 Flash; $+$0.10 pp for Gemini 3 Flash), suggesting their visual analysis adds little beyond what the captions already convey. In contrast, GPT-5-mini degrades sharply ($+$9.55 pp: 18.41\% vs.\ 8.86\%) and Qwen degrades moderately ($+$3.29 pp: 22.93\% vs.\ 19.64\%), indicating that direct visual reasoning provides substantially richer information for these models. For damage severity, the pattern reverses for GPT-5-mini: Text+Caption actually outperforms Multimodal ($+$2.98 pp improvement: 11.49\% vs.\ 14.47\%), consistent with the finding that raw post-recession imagery introduces misleading signals for GPT-5-mini's damage estimates, whereas textual captions appear to filter this noise. Across Gemini and Qwen, damage MAE is similar between the two configurations (within 1.2 pp). Taken together, these results suggest that caption-based substitution is viable for models with strong text-based flood reasoning (Gemini family), while models that benefit most from raw imagery (GPT-5-mini for extent) cannot be adequately served by text descriptions alone.

\medskip
\textbf{(3) Social media}
\medskip

Removing tweets while retaining imagery degrades flood extent consistently across all four models relative to the full multimodal baseline (Gemini 2.5 Flash: from 12.98\% to 17.42\%; Gemini 3 Flash: from 11.11\% to 17.70\%; Qwen: from 19.64\% to 27.10\%; GPT-5-mini: from 8.86\% to 14.68\%), with degradations of $+$4.44 to $+$7.46 pp, confirming that social media is the primary real-time signal for hazard estimation. Damage MAE is largely stable without tweets for Gemini variants, with only modest degradation for Qwen ($+$1.31 pp) and GPT-5-mini ($+$0.92 pp). This asymmetry reflects the specialized roles of each text source: social media captures dynamic flooding events (rising water levels, road closures, rescue requests) that are the primary real-time evidence for extent estimation. Structural damage, by contrast, is signaled more directly through 311 emergency calls, which log property-level flooding complaints, and through historical damage patterns encoded in the FEMA knowledge base. These sources remain intact when tweets are removed, which is why damage prediction is largely preserved.

\subsection{Supplementary Analysis: Full 207-ZIP Text-Only Results}
\label{sec:suppl_207}

\begin{figure}[!htbp]
    \centering
    \includegraphics[width=0.72\textwidth]{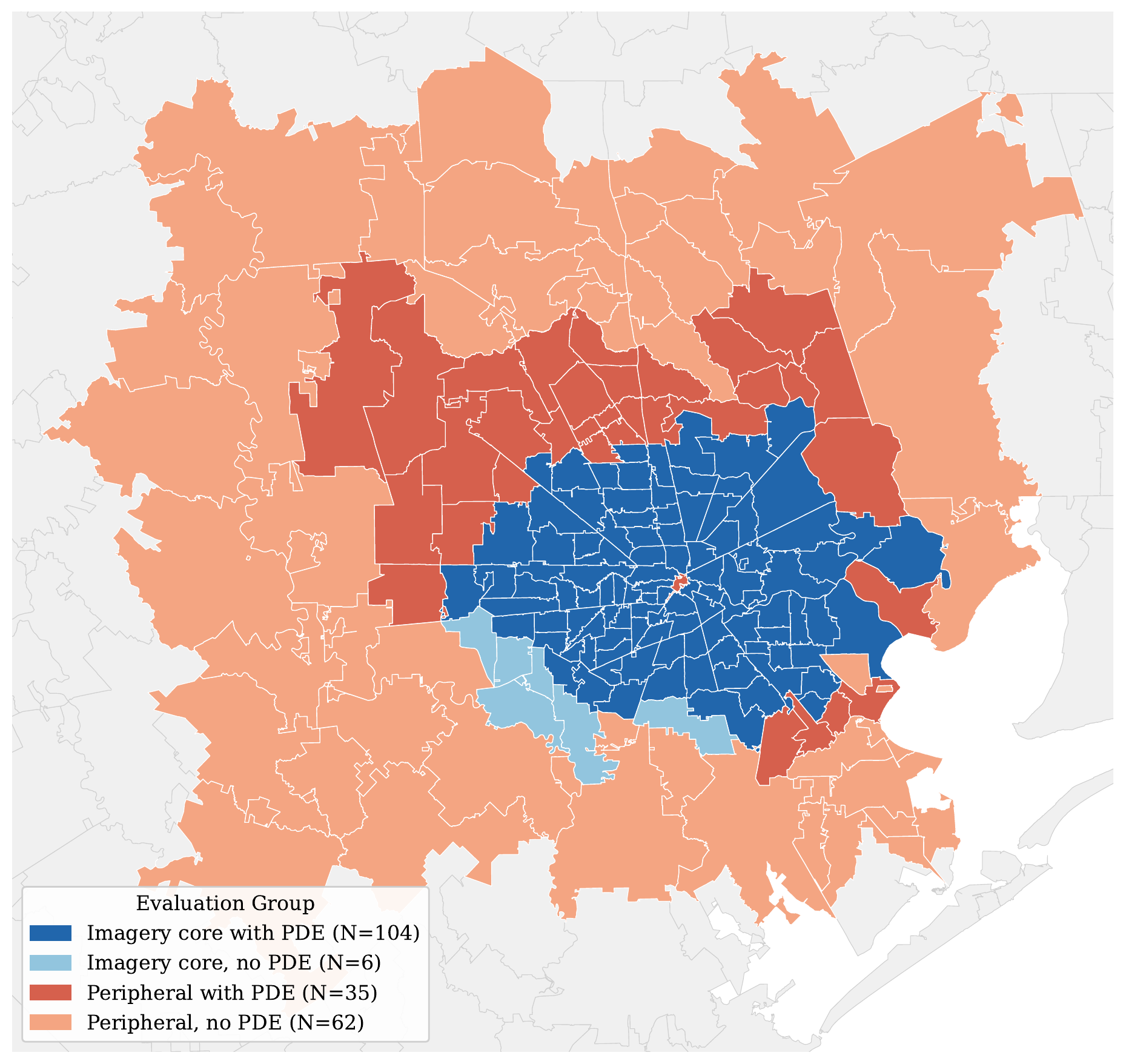}
    \caption{Evaluation group coverage across the 207-ZIP study area. Blue: 110-ZIP imagery core (dark = PDE available, $N=104$; light = no PDE, $N=6$). Red: 97 peripheral ZIPs for text-only supplementary analysis (dark = PDE available, $N=35$; light = no PDE, $N=62$). The imagery core covers the urban center and inner suburbs; peripheral ZIPs ring the outer boundary. A few small central ZIPs (e.g., downtown 77002) appear red despite their central location due to insufficient aerial tile coverage.}
    \label{fig:coverage_groups}
\end{figure}

As described in Section~\ref{sec:dataset}, the main ablation results are restricted to the 110 imagery-covered ZIPs. Here we report text-only performance on the remaining 97 peripheral ZIPs and on the combined 207-ZIP study area, providing a broader picture of how the text-only pipeline generalizes beyond the imagery-covered core. Figure~\ref{fig:coverage_groups} maps the spatial distribution of both groups.

\begin{table}[!htbp]
    \centering
    \caption{Text-Only RAG performance on the 97 ZIPs without imagery coverage and on the full 207-ZIP study area (bootstrapped 95\% CI). Flood extent MAE: $N=97$ / $N=207$; Damage MAE: $N=35$ / $N=139$ (ZIPs with PDE ground truth only).}
    \label{tab:suppl_207}
    \resizebox{\textwidth}{!}{%
    \begin{tabular}{lcccc}
    \toprule
    & \multicolumn{2}{c}{\textbf{No-Imagery (Peripheral) ZIPs}} & \multicolumn{2}{c}{\textbf{Full Study Area}} \\
    \cmidrule(lr){2-3} \cmidrule(lr){4-5}
    \textbf{Model} & \textbf{Ext MAE ($N=97$)} & \textbf{Dmg MAE ($N=35$)} & \textbf{Ext MAE ($N=207$)} & \textbf{Dmg MAE ($N=139$)} \\
    \midrule
    Gemini 2.5 Flash & 14.97 (12.78--17.24) & 10.92 (8.07--14.11) & 13.86 (12.50--15.28) & 11.64 (10.05--13.25) \\
    Gemini 3 Flash   & 14.42 (12.54--16.41) & 11.76 (9.20--14.53) & 12.74 (11.48--14.06) & 10.47 (9.17--11.79) \\
    Qwen 3.5         & 20.53 (18.45--22.72) & 14.12 (11.78--16.56) & 21.34 (19.94--22.87) & 12.44 (11.03--13.86) \\
    GPT-5-mini       & 16.87 (14.62--19.30) & 13.26 (10.50--16.15) & 17.19 (15.73--18.62) & 11.67 (10.24--13.18) \\
    \bottomrule
    \end{tabular}%
    }
\end{table}

Two patterns emerge. First, flood extent MAE on the no-imagery (peripheral area) ZIPs is broadly consistent with the 110-ZIP (core area) text-only baseline (Table~\ref{tab:ablation_configs}), showing that the text retrieval pipeline generalizes across the full study area without degradation. Second, text-only damage MAE on the no-imagery ZIPs is of similar magnitude but somewhat more variable than on the imagery-covered set (10.92--14.12\% vs.\ 11.14--11.88\%), reflecting the smaller damage-evaluation sample and the sparser evidence environment in peripheral areas. The no-imagery ZIPs have substantially lower PDE coverage ($N_\text{dmg}=35$ out of 97) and represent more peripheral areas where both social media evidence and structural damage signals are weaker.

\subsection{Error Analysis}
\label{sec:error_analysis}

We present two representative failure cases drawn from Gemini~2.5 Flash multimodal runs that illustrate the primary error modes of our system.

\noindent\textbf{Failure Case 1}: Underestimation (ZIP 77026).
Ground Truth: 72.3\% | Prediction: 35.0\% | Abs.\ Error: 37.3 pp

{\sloppy
\begin{description}
    \raggedright
    \item[Retrieved Evidence:]
    Tweets (20): ``Flooding on 610 N Lp''; ``Closed due to flooding'' on I-69 Eastex Fwy and its Frontage Road.
    Sensors: 3 HCFCD gauges retrieved.
    311 Calls: 20 (multiple ``damage by storm or flood'' reports).
    Imagery: 6 tiles retrieved; visual analysis failed (returned 0\%).

    \item[Model Reasoning:] ``ZIP 77026 experienced significant flooding\ldots evidenced by multiple ZIP-specific tweets reporting `Flooding' and `Closed due to flooding' on major roadways\ldots While specific water depths or detailed damage descriptions are not available, the combination of extensive road closures and multiple property damage reports suggests a substantial portion of the ZIP was impacted\ldots''

    \item[Failure Mode:] Road-corridor flooding reports (highway closures on linear infrastructure) do not capture how far floodwater penetrated residential neighborhoods away from the corridors. The model correctly inferred flooding presence but anchored its spatial extent estimate on road-closure footprints rather than area-wide inundation. Visual analysis also failed entirely for this ZIP, removing an important confirmatory signal; the fusion engine fell back to the conservative text-only estimate of 35\%, well below the 72.3\% ground truth.
\end{description}
}

\noindent\textbf{Failure Case 2}: Overestimation (ZIP 77056).
Ground Truth: 37.7\% | Prediction: 70.0\% | Abs.\ Error: 32.3 pp

{\sloppy
\begin{description}
    \item[Retrieved Evidence:] \mbox{}\\
    Tweets (20): ``road, er\ldots I mean river getting out of our property'' (901830606081802240); ``Westpark Tollway closed due to flooding'' (901765948750917632); ``Buffalo Bayou `so far over its banks' near 610 West'' (902224515702415365).
    Sensors: 3 HCFCD gauges.
    311 Calls: 20 (multiple Flooding and Street Hazard entries).
    Imagery: 6 tiles; visual analysis failed (returned 0\%).

    \item[Model Reasoning:] ``Multiple tweets confirm widespread street flooding\ldots Buffalo Bayou was `so far over its banks'\ldots Given the descriptions of streets turning into rivers and major bayou overflow, it is highly probable that water entered some ground-level properties\ldots The widespread nature of street and bayou flooding indicates a large portion of the ZIP was affected.''

    \item[Failure Mode:] Vivid, high-salience language (``road\ldots I mean river'', ``so far over its banks'') and landmark events (Buffalo Bayou overflow, major highway closure) were extrapolated to ZIP-wide inundation. ZIP 77056 encompasses the Galleria/Uptown district, which is a largely commercial and topographically elevated area where bayou flooding near the western edge does not translate to 70\% areal inundation. The model over-weighted the severity of localized landmark events rather than reasoning about the spatial distribution of flood risk within the ZIP boundary.
\end{description}
}

Our analysis reveals four primary error categories: (1) corridor-anchored underestimation, where road-closure and linear-infrastructure reports cause the model to underestimate area-wide inundation; (2) salience extrapolation, where vivid landmark flood events inflate ZIP-wide estimates beyond localized impacts; (3) false positives from misinterpreting normal water features as flood damage; and (4) data sparsity, where ZIP codes lacking local social media or 311 reports receive near-zero predictions despite actual flooding. Qualitative success cases showing accurate regional and local inference are provided in Appendix~\ref{sec:qualitative_examples}.

% CONCLUSION
%==============================================================================
\section{Conclusion}
\label{sec:conclusion}

We presented CrisiSense-RAG, a multimodal RAG framework for disaster impact assessment that deploys general-purpose pre-trained models in a zero-shot setting, without disaster-specific fine-tuning. Evaluated on Hurricane Harvey across four model backends, the full multimodal configuration achieves Extent MAE of 8.86--19.64\% and Damage MAE of 10.14--14.47\%, with the largest flood extent improvement over text-only inference reaching 8.62 percentage points.

The split-pipeline architecture is central to these results: by routing text and visual evidence through separate analysts before temporally-aware fusion, the system prevents post-event imagery from negating real-time flood reports while allowing imagery to contribute when it improves flood extent estimation. Metric-aligned prompting further ensures outputs conform to operational ground-truth definitions rather than qualitative severity language. Together, these design choices establish that carefully structured zero-shot inference is a practical and immediately deployable baseline for rapid disaster response, accessible without the resource-intensive requirements of collecting large task-specific datasets.

\subsection{Limitations and Future Work}

\textbf{Uncertainty quantification} A key limitation of the current system is that it produces point estimates with no calibrated uncertainty bounds. The \texttt{confidence} field in the output is self-reported by the LLM and is not statistically grounded. In practice, predictions backed by 20 location-specific tweets and 311 calls should carry more weight than those backed by a handful of generic regional reports. The fusion engine already records source conflicts (cases where text and visual estimates disagree), which could serve as a proxy for prediction uncertainty. A natural next step would be to derive prediction intervals, for example, by running the pipeline multiple times with bootstrap-sampled subsets of retrieved documents, or by ensembling across model providers and measuring the spread. Until such calibration is in place, we recommend that operational users treat predictions with low evidence counts or high source disagreement as requiring manual review.

\textbf{Generalizability} Our evaluation focuses on a single event (Hurricane Harvey, Harris County, TX), and applying the system to a different disaster would require some adaptation. To be specific about what transfers and what does not: the RAG pipeline architecture, fusion logic, retrieval models, and the general structure of the prompts are all disaster-domain general and would carry over without changes. The FEMA historical priors cover all US ZIP codes with declared disasters since 2010, so they apply nationwide. What would need updating for a new event is the data ingestion layer: raw data sources, time windows, geographic filters, and any event-specific metadata (e.g., peak dates for the temporal reasoning prompts). This adaptation requires modest engineering effort (updating filters and metadata), not model retraining. That said, generalizability to non-flood hazards (wildfires, earthquakes) and non-US geographies remains unverified and is a clear direction for future work.

\textbf{Other limitations} Semantic retrieval improves coverage over coordinate-based filtering but may introduce spatial noise when tweets about one neighborhood get retrieved for a neighboring ZIP code. The damage evaluation is also constrained by the availability of PDE ground truth for only 139 of the 207 ZIP codes.

Future work should extend the framework to multi-hazard events, incorporate real-time streaming data sources, and add the calibrated uncertainty estimates discussed above. Developing human-in-the-loop interfaces for expert validation would further improve both accuracy and operational trust.

\subsection{Broader Impact}
This framework has the potential to assist emergency responders in rapid damage assessment by synthesizing heterogeneous information sources. Unlike black-box models, the retrieval-based design provides an auditable trail of evidence (e.g., source tweets, imagery), allowing analysts to verify predictions. However, automated systems should augment rather than replace human judgment. Our error analysis reveals susceptibility to contrarian reports and regional over-generalization, which could misallocate resources if predictions are accepted uncritically. We recommend operational deployments include confidence thresholds for human review and explicit reasoning disclosure for verification. Finally, communities with lower social media penetration may be underrepresented, and future work should evaluate and mitigate such biases to ensure equitable disaster response.

%==============================================================================
% ACKNOWLEDGEMENTS
%==============================================================================
\section*{Acknowledgements}
The authors would like to thank Junwei Ma for providing the Property Damage Extent (PDE) dataset used as part of the ground truth in this study.

%==============================================================================
% FUNDING
%==============================================================================
\section*{Funding}
This research did not receive any specific grant from funding agencies in the public, commercial, or not-for-profit sectors.

%==============================================================================
% CODE AND DATA AVAILABILITY
%==============================================================================
\section*{Code and Data Availability}
The code used in this study is publicly available on \href{https://github.com/YimingXiao98/CrisiSense-RAG}{GitHub}. The dataset is available on \href{https://huggingface.co/datasets/Ymx1025/CrisiSense-RAG-dataset}{Hugging Face}.

%==============================================================================
% REFERENCES
%==============================================================================
\bibliography{references}
\bibliographystyle{unsrtnat}

%==============================================================================
% APPENDIX
%==============================================================================
\newpage
\appendix

\section{Prompts}
\label{sec:prompts}

We provide the full prompts used in our experiments. The temporal context and semantic alignment instructions (highlighted in the system prompt) were critical for achieving the best performance.

\subsection{System Prompt (Multimodal)}
\label{sec:system_prompt}

The following system prompt is used for both Text+Caption and full Multimodal experiments:

\begin{small}
\begin{verbatim}
You are an assistant estimating post-disaster impact for Harris County,
TX during Hurricane Harvey. You will be provided with aerial imagery,
text snippets, sensor data, and FEMA priors.

## CRITICAL TEMPORAL CONTEXT:
- Hurricane Harvey PEAK FLOODING: August 27-28, 2017
- Aerial imagery captured: August 31, 2017 (3-4 days AFTER peak)
- By Aug 31, most floodwaters had RECEDED from streets
- Text reports (tweets/311 calls) are from DURING the event (real-time)

## KEY INSIGHT:
If imagery shows "dry/clear" but text reports "flooded", the flooding
DID happen - the water simply receded before the aerial imagery captured
the image. TRUST THE TEXT.

IMPORTANT: All provided text snippets (Tweets, 311 Calls) are
pre-filtered and RELEVANT to the queried location/time.

## DECISION RULES:
- For FLOOD EXTENT: PRIORITIZE text reports. Visual "no flooding"
  means water receded, NOT that it didn't flood.
- For STRUCTURAL DAMAGE: Visual IS reliable for persistent damage
  (debris, destroyed buildings). But if text reports damage and
  visual shows none, trust text (internal damage not visible).

## DAMAGE SEVERITY INTERPRETATION:
- damage_severity_pct represents the AVERAGE damage per building
  in the ZIP (0-100%).
- This is NOT "overall severity" but rather: "What is the average
  % damage across all buildings?"
- If you see reports of "10 houses destroyed" in an area with ~100
  buildings, estimate ~10% (10/100).
- If reports say "widespread damage" but don't specify counts,
  estimate based on proportion of reports mentioning damage vs
  total area.

## CHAIN OF THOUGHT REASONING:
- In "reasoning", list every Tweet ID and 311 Call ID you see.
- Note any temporal discrepancy between text and imagery.
- For flood extent, base your estimate primarily on text evidence.

## EXAMPLE:
Input Context:
- Images: [IMG_1] (shows clear roads - captured Aug 31)
- Tweets: [T123] (ZIP 77002, Aug 27) "Water entering my living room!"

Correct Output Reasoning:
"Tweet T123 reports water in living room on Aug 27. Image IMG_1
from Aug 31 shows clear roads. This is expected - water receded
by Aug 31. I estimate HIGH flood extent based on the tweet."

Respond with valid JSON matching the schema provided by the user.
\end{verbatim}
\end{small}

\subsection{Text-Only System Prompt}
\label{sec:text_only_prompt}

For the Text-Only baseline, we use a distinct prompt focused on evidence synthesis from text and rainfall signals, without visual references:

\begin{small}
\begin{verbatim}
You are an assistant estimating post-disaster impact for Harris County,
TX during Hurricane Harvey (Aug 25-Sep 1, 2017). You will be provided
with tweets, 311 calls, peak rainfall observations, and a historical
flood risk prior.

## EVIDENCE HIERARCHY -- use in this order when estimating:
1. ZIP-specific tweets/311 calls (highest weight -- direct real-time
   local reports)
2. Peak rainfall inches -- direct observational evidence; use as flood
   extent floor
3. Regional Harvey tweets ("flooding in Houston area") -- confirms
   county-wide flooding
4. Historical NFIP risk profile -- vulnerability prior only, NOT
   Harvey-specific damage

## CRITICAL: REGIONAL HARVEY TWEETS ARE LOCAL EVIDENCE
During Harvey, Harris County experienced catastrophic uniform flooding
across the metro area. Tweets mentioning "flooding in Houston area",
"Harvey flooding", "1000-year flood event", "historic flooding in
Houston" ARE evidence that flooding occurred in every Harris County
ZIP. Do NOT dismiss these as "too general." Their presence means
flood_extent_pct >= 15% for any Harris County ZIP. Cite them as
supporting evidence in your reasoning.

## NEVER RETURN 0% WHEN HARVEY EVIDENCE EXISTS
Do NOT return flood_extent_pct=0 or damage_severity_pct=0 if ANY of
the following are true:
- Peak rainfall >= 20 inches for this ZIP
- Any tweet mentions Harvey flooding in Houston/Harris County
In these cases, the ZIP was affected -- estimate the degree, not
whether it was affected.

## DAMAGE SEVERITY:
- damage_severity_pct = average fraction of building value destroyed
  across all buildings (0-100%).
- Water depth signals: <1 ft=minor (5-15%), 1-3 ft=moderate (15-35%),
  3-6 ft=severe (35-60%), >6 ft=catastrophic (60-100%).
- Tweet language: "water in living room" (~15-30%),
  "total loss/destroyed" (~70-100%), "flooded but OK" (~5-20%).
- If tweet/311 evidence is sparse, use rainfall magnitude as proxy
  for damage floor.

Respond with valid JSON matching the schema provided by the user.
\end{verbatim}
\end{small}

\subsection{User Prompt Template}
\label{sec:user_prompt}

Each query provides rainfall observations, a historical NFIP risk prior, retrieved tweets and 311 calls, and (for multimodal configurations) image captions with temporal warnings:

\begin{small}
\begin{verbatim}
ZIP: {zip_code}
Time window: {start} to {end}
Imagery IDs: [tile_id_1, tile_id_2, ...]

### Rainfall Observations (Direct Evidence -- HCFCD gauge network,
near-real-time):
{peak_rainfall_inches and spatial coverage statistics for this ZIP}

### Historical Flood Risk Prior for ZIP {zip_code}
(Pre-event vulnerability context only):
{NFIP claim history and historical loss statistics}

### Tweets (Relevant to ZIP {zip_code}):
NOTE: Tweets tagged "(Content Relevant)" are regional Harvey reports
-- they confirm flooding occurred across Harris County including this
ZIP. Do NOT dismiss them as too general.
- [tweet_id] (Verified ZIP {zip_code}) {tweet_text}
- [tweet_id] (Content Relevant) {tweet_text}
...

### 311 Calls (ZIP {zip_code}, Exact Match):
- [call_id] (Verified ZIP {zip_code}) {call_description}
...

### Image Captions  [text+caption configuration only]
(TEMPORAL WARNING):
These captions describe imagery from Aug 31, 2017 - AFTER peak
flooding. By Aug 31, floodwaters had RECEDED. "No flooding visible"
does NOT mean flooding didn't occur!
- For FLOOD EXTENT: IGNORE these captions. Trust tweets/311 calls.
- For STRUCTURAL DAMAGE: These captions ARE useful (damage persists).
- [tile_id] {caption_text}
...

Respond with JSON matching schema:
{
  "reasoning": str,
  "zip": str,
  "time_window": {"start": str, "end": str},
  "estimates": {
    "flood_extent_pct": float,    // % of ZIP area flooded (0-100)
    "damage_severity_pct": float, // Avg structural damage per building
                                  // (0-100). MEAN across all buildings.
    "roads_impacted": list[str],
    "confidence": float           // 0-1
  },
  "evidence_refs": {
    "imagery_tile_ids": list[str],
    "tweet_ids": list[str],
    "call_311_ids": list[str]
  },
  "natural_language_summary": str
}
\end{verbatim}
\end{small}

\subsection{Query Parsing Prompt}
\label{sec:query_parsing_prompt}

For the natural language chat interface, we use a query parsing prompt to extract structured parameters from user requests:

\begin{small}
\begin{verbatim}
You are a query parser for a disaster impact assessment system.
Your goal is to extract structured parameters from a natural language 
user request. The user is asking about Hurricane Harvey impact in a 
specific location and time.

Extract the following fields:
- zip: The 5-digit ZIP code (e.g., "77096"). If missing, return null.
- start: The start date in YYYY-MM-DD format.
- end: The end date in YYYY-MM-DD format.

If only one date is mentioned, use it for both start and end.
If no year is mentioned but "Harvey" is implied, assume 2017.
If no date is mentioned, return null for dates.

Respond with valid JSON only.
\end{verbatim}
\end{small}

The user message is then formatted as:

\begin{small}
\begin{verbatim}
User Message: "{message}"

Respond with JSON matching schema:
{
  "zip": str | null,
  "start": str | null,
  "end": str | null
}
\end{verbatim}
\end{small}

\section{Tweet Filtering Keywords}
\label{sec:tweet_keywords}

We implemented a keyword-based filtering pipeline to improve the signal-to-noise ratio of the $\sim$24 million raw tweets. A tweet is included if it contains at least one allow keyword and no block keywords.

\subsection{Allow List}
\label{sec:allow_list}

Keywords that indicate disaster-relevant content:

\begin{center}
\begin{tabular}{llll}
\toprule
flood & flooding & flooded & hurricane \\
storm & rain & underwater & rescue \\
trapped & stuck & help & emergency \\
911 & evacuate & damage & collapsed \\
power & outage & road & bridge \\
bayou & creek & & \\
\bottomrule
\end{tabular}
\end{center}

\subsection{Block List}
\label{sec:block_list}

During initial corpus exploration, we observed that many tweets containing disaster-related terms (e.g., Harvey, Houston) were unrelated to the hurricane. Common sources of noise included: (1) music promotion and streaming service spam using trending hashtags; (2) political commentary co-opting the disaster for unrelated messaging; (3) commercial advertisements and promotional giveaways; and (4) sports and entertainment discussions. The block list was developed iteratively by examining false positives in the filtered corpus.

Keywords that indicate irrelevant content:

\begin{center}
\begin{tabular}{llll}
\toprule
spotify & music & song & album \\
lyrics & vote & election & trump \\
biden & president & giveaway & contest \\
win & sale & shirt & merch \\
game & nfl & nba & football \\
baseball & love & heart & tears \\
\bottomrule
\end{tabular}
\end{center}

This filtering reduced the corpus from $\sim$24 million to 458,453 tweets ($\sim$1.9\% acceptance rate).

\section{Successful Examples}
\label{sec:qualitative_examples}

We present two success cases drawn from current model runs demonstrating accurate flood extent prediction when evidence is convergent and correctly interpreted.

\noindent\textbf{Success Case 1}: Convergent Multimodal Fusion (ZIP 77040, GPT-5-mini).
Ground Truth: 45.1\% | Prediction: 45.1\% | Abs.\ Error: 0.0 pp

{\sloppy
\begin{description}
    \item[Retrieved Evidence:]
    Tweets (20): local reports of White Oak Bayou overbanking, US-290 NW Freeway flooding at Gessner (901043725001949185);\allowbreak{} ``water rose and later fell'' reports;\allowbreak{} ``water didn't settle here'' (901661375474360320).
    Sensors: 3 HCFCD gauges.
    311 Calls: 20 (Drainage, Flooding, storm-debris, and wastewater requests).
    Imagery: 6 tiles; sediment and debris deposits confirming past flooding, some roof damage visible.

    \item[Model Reasoning:] ``The mix of evidence indicates neighborhood/stream overbank flooding affecting a substantial fraction of the ZIP but not wholesale catastrophic inundation: localized bayou overbanking and multiple drainage/flooding complaints support a moderate flood footprint. Tweet language does not describe wholesale or total-loss/structural collapse; messages like `water has gone down a little' imply typical Harvey-era depths in many residential locations in the 1--3~ft range but with many properties less affected.''

    \item[Why It Worked:] All evidence streams were internally consistent. Text analysis (35\%) was conservative, anchored by calibrated tweet language. Aerial imagery independently confirmed flooding (60\%). The temporal-aware fusion engine blended the two signals to 45.1\%, matching ground truth exactly. No conflict resolution was triggered, allowing the fusion weights to operate as designed.
\end{description}
}

\noindent\textbf{Success Case 2}: Robust Text-Primary Inference (ZIP 77020, Gemini~3 Flash).
Ground Truth: 70.4\% | Prediction: 70.0\% | Abs.\ Error: 0.4 pp

{\sloppy
\begin{description}
    \item[Retrieved Evidence:]
    Tweets (20): multiple reports of total I-10 (Baytown East Fwy) closure at Lockwood, Waco, Wayside, McCarty, and Gregg (901661136\allowbreak{}583577600,\allowbreak{} 901793077\allowbreak{}983150081);\allowbreak{} post-storm ``salvaging,'' ``throwing out,'' and ``cleaning up'' reports (903605794\allowbreak{}981036034,\allowbreak{} 903616888\allowbreak{}026021890). % chktex 38
    Sensors: 3 HCFCD gauges.
    311 Calls: 20 (Medical Evacuation, ``Document damage by storm or flood'' entries).
    Imagery: 6 tiles; visual analysis failed (returned 0\%).

    \item[Model Reasoning:] ``The total closure of I-10 at multiple major intersections indicates the primary east-west artery through the ZIP was impassable for several days. Post-storm tweets explicitly mention `salvaging,' `throwing out,' and `cleaning up,' confirming that floodwaters entered residential or commercial structures. 311 calls for Medical Evacuation and storm damage further support a high impact.''

    \item[Why It Worked:] Visual analysis failed entirely for this ZIP, yet the prediction was near-perfect. The text-primary architecture preserved accuracy: the model correctly weighted structural cleanup language (``salvaging,'' ``throwing out'') as strong evidence of residential inundation, independent of imagery. This demonstrates the resilience of the pipeline when one modality is unavailable.
\end{description}
}

\end{document}